\documentclass[fleqn,usenatbib]{mnras}

\usepackage{newtxtext,newtxmath}
\usepackage{mathptmx}
\usepackage{txfonts}
\usepackage{float}
\usepackage{amsmath}	% Advanced maths commands
\usepackage{amssymb}
\usepackage[T1]{fontenc}

\DeclareRobustCommand{\VAN}[3]{#2}
\let\VANthebibliography\thebibliography
\def\thebibliography{\DeclareRobustCommand{\VAN}[3]{##3}\VANthebibliography}

\usepackage{graphicx}	% Including figure files
\usepackage{amsmath}	% Advanced maths commands
\usepackage{amssymb}	% Extra maths symbols

\def\la{\raise.5ex\hbox{$<$}\kern-.8em\lower 1mm\hbox{$\sim$}}
\def\ma{\raise.5ex\hbox{$>$}\kern-.8em\lower 1mm\hbox{$\sim$}}

\def\kms{${\rm km}\,{\rm s}^{-1}$}
\def\cm3{cm$^{-3}$}

\def\Vs{$V_{\rm s}$}
\def\n0{$n_{\rm 0}$}
\def\B0{$B_{\rm 0}$}

\def\Ne{$n_{e}$}
\def\Te{$T_{\rm e}$}

\def\erg{$\rm erg\, cm^{-2}\, s^{-1}$}

\def\L12{$L_{12\mu m}$}
\def\F12{$F_{12\mu m}$}

\def\Hb{H${\beta}$}

\def\Ha{H${\alpha}$}

\def\o3{[O\,{\sc iii}]\,$\lambda$5007}
\def\fe7{[Fe\,{\sc vii}]\,$\lambda$6087}

\def\outs{outflows} 
\def\ff{{\it ff}}

\title[The CLR in IC\,5063]{Physical Conditions and Extension of the Coronal Line Region in IC\,5063}

\author[M. A. Fonseca-Faria et al.]{
M. A. Fonseca-Faria,$^{1,2}$\thanks{E-mail: marcosfonsecafaria@gmail.com}
A. Rodr\'{\i}guez-Ardila,$^{1,2}$
M. Contini,$^{3}$
L. G. Dahmer-Hahn,$^{4,1}$
R. Morganti$^{5,6}$
\\
$^{1}$Laborat\'orio Nacional de Astrof\'{\i}sica, Rua dos Estados Unidos, 154, Itajub\'a, MG, Brazil\\
$^{2}$Instituto Nacional de Pesquisas Espaciais, Av. dos Astronautas, 1758 - Jardim da Granja S\~ao Jos\'e dos Campos/SP - CEP 12227-010, Brazil\\
$^{3}$School of Physics and Astronomy, Tel Aviv University, Tel Aviv 69978, Israel\\
$^{4}$Shanghai Astronomical Observatory, Chinese Academy of Sciences, 80 Nandan Road, Shanghai 200030, China\\
$^{5}$ Kapteyn Astronomical Institute, University of Groningen, P.O. Box 800,
9700 AV Groningen, The Netherlands. \\
$^{6}$ ASTRON, the Netherlands Institute for Radio Astronomy, Oude Hoogeveensedijk 4, 7991 PD Dwingeloo, The Netherlands.
}

\date{Accepted XXX. Received YYY; in original form ZZZ}

\pubyear{2015}

\begin{document}
\label{firstpage}
\pagerange{\pageref{firstpage}--\pageref{lastpage}}
\maketitle

\begin{abstract}
We study the ionised and highly ionised gas phases in the Seyfert~2 galaxy IC~5063 by means of VLT/MUSE integral field spectroscopy. Our analysis allowed us to detect a high-ionisation gas outflow traced by the coronal lines [\ion{Fe}{vii}]~$\lambda$6087 and [\ion{Fe}{x}]~$\lambda$6375. Both emissions are found to be extended. The former up to 1.2~kpc and 700~pc NW and SE from the nucleus, respectively. The latter reaches 700~pc NW of the nucleus. This is the first time that [\ion{Fe}{x}] emission is observed at such distances from the central engine in an active galactic nucleus. The [\ion{Fe}{vii}]~$\lambda$6087 emission peaks at the nucleus, with two secondary peaks at the position of the NW and SE radio-lobes. The gas kinematics is complex, with the coronal emission displaying split line profiles along the radio jet and line widths of several hundreds km~s$^{-1}$. Velocity shifts of up to 600~km~s$^{-1}$ in excess of the systemic velocity of the galaxy are found very close to the radio lobes and along the jet propagation. The extended coronal gas is characterised by temperatures reaching 20000~K and electron densities $>10^2$~cm$^{-3}$, with the larger values associated to the regions of larger turbulence, likely due to the passage of the radio jet. This hypothesis is supported by photoionisation models that combine the effects of the central engine and shocks. Our work highlights the strong relationship between extended coronal emission and the radio jet, with the former suitably tracing the latter, which in the case of IC~5063, propagates very close to the galaxy disc.

\end{abstract}


%%%%%%%%%%%%%%%%%%%%%%%%%%%%%%%%%%%%%%%%%%%%%%%%%%

%%%%%%%%%%%%%%%%% BODY OF PAPER %%%%%%%%%%%%%%%%%%

\section{Introduction}

Feedback from active galactic nuclei (AGN) is 
an important driver of galaxy evolution \citep{fabian_2012,silk_1998}. 
Currently, cosmological models of galaxy formation adopt this prescription
in order to match observational data \citep{granato_2004, springel_2005, croton_2006,lapi_2006, lapi_2014}.
Two main feedback channels have now been identified. 
The radio- or kinetic-mode, and the quasar- or radiative-
mode. The former is usually associated with powerful radio sources
with low accretion efficiencies. In these AGN,
large scale relativistic jets powered by the
central black hole can
transfer energy to its environment, preventing the gas from cooling and, therefore, affecting the stellar mass
assembly of galaxies \citep{mcnamara_2017,fabian_2012}.
The quasar radiative mode is expected to be dominant at higher accretion 
efficiencies, where AGN-driven winds of ionized, neutral, and molecular 
gas are produced by radiation pressure on the circumnuclear gas \citep{fabian_2012}.

Although the feedback scenario in powerful radio-sources or in highly luminous 
quasars is firmly established, the question over what is the
dominant feedback process in moderate- to low-luminosity AGN with modest
radio-jets is rather uncertain \citep{wylezalek_2018}. 
This is mainly because moderate radio jets are ubiquitous in local
Seyfert galaxies and these jets can interact with the
ISM gas on galactic scales \citep{morganti_1999,thean_2000}. While the complex nature of the interaction is
still under study, there is evidence that these low-power radio jets can produce
galactic winds which, in turn, interact with dense multiphase
gas \citep[see][and references
therein]{rosario_2018,riffel_2014b}. Thus, relevance of the kinetic channel as a major way of
releasing nuclear energy to the ISM in low-luminosity active
galactic nuclei (AGN) has been underestimated so far \citep{merloni_2013,wylezalek_2018}. 

Indeed, recent works reveal that low-power radio jets can
play a major role in driving fast, multiphase, galaxy-scale
outﬂows in radio-weak AGN \citep{ardila_2017,may_2018,fabbiano_2018a,ardila_2020,venturi_2022,murthy_2022}. 
Mass outﬂow rates,
of up to 8~M$_{\odot}$~yr$^{-1}$, similar to those found in powerful 
radio-loud AGN have been derived. In some of these works, high-ionisation lines (or coronal lines, CLs) such as [\ion{Fe}{vii}]~$\lambda$6087 in the optical or [\ion{Si}{vi}]~$\lambda$19641 in the NIR  have been employed 
to trace the most energetic component of AGN feedback. It is important to note that high-ionisation lines are not produced by stars, being more reliable indicators for AGN-driven processes. Morevover, coronal lines in general (IP $\geq$ 100~eV) may have a physical connection with the X-ray emission in terms of location and physical conditions. Very recently, \citet{falcao_2022} showed via photoionisation modeling that CLs from ions with ionization potential greater than or equal to that of \ion{O}{vii}, i.e. 138 eV trace the footprint of X-ray gas. Therefore, they can be used to measure the kinematics of the extended X-ray emitting gas at the spectral resolution dictated by optical spectrographs. In this respect,  coronal lines allow us to detect the high-, and the highest-excitation component of the outflow in AGN.

An example of this is the Circinus galaxy where \citet{ardila_2020} revealed that the [\ion{Fe}{vii}]
emission extends up to a distance of 700~pc from the AGN. The coronal gas extension 
is likely the remnant of shells inﬂated by the passage of a radio jet. This
scenario is supported by \citet{fonseca-faria_2021} by means of
models that combine the effects of shocks driven by the jet and photoionisation by the central source. Similar results showing the strong relationship between 
the jet and extended coronal emission were recently presented by 
\citet{speranza_2022}. Overall, the above findings all point towards the
relevance of low-accretting sources in the delivery of kinetic
energy to the ISM. They are also in agreement to simulations
of radio jets emerging out of the host’s ISM which show that the jets vigorously
distort the gas distribution in the gaseous galactic discs \citep{gaibler_2012,cielo_2018,mukherjee_2018,Mukherjee_2018b}.
The coupling between the jet and the ISM strongly
depends on the jet’s inclination with respect
to the galactic disc \citep{Mukherjee_2018b}. 

In this context, IC~5063 (RA=20:52:02, DEC=-57:04:07; J2000) is a prime target to study the relationship
between the radio-jet and the coronal emission. It is a Seyfert~2 galaxy \citep[][and references therein]{Colina91} that hosts a low radio power jet 
(P$_{1.4}$~GHz~$\sim$10$^{23.4}$~W~Hz$^{-1}$) and an optical/NIR spectrum with strong emission lines, including coronal lines \citep{Colina91}.  Despite the low radio power, it provides one of the clearest examples of jet-induced
outflows, where the radio plasma is disturbing the kinematics of all the phases of the gas \citep[see][for an overview]{tadhunter_2014,morganti_2015}.  The region co-spatial
with the radio emission is where the most kinematically disturbed ionised, molecular and
H I gas is located (with velocities deviating up to 600~km~s$^{-1}$ from regular rotation). This
jet-ISM interaction is also affecting the physical conditions of the gas \citep{oosterloo_2017}. The properties observed have been well reproduced by hydrodynamic simulations
and the details of the comparison between the data and the simulation is presented in
\citet{mukherjee_2018}. As seen from above, IC\,5063  has been studied in detailed in multiple bands and we will describe some of those results later in the text, comparing them to our findings.

Here, we report optical observations of IC~5063
focusing on three main aspects: (i) the determination of the full
extension of the coronal gas and its relationship with galactic
feedback; (ii) the role of a jet in shaping the morphology and kinematics of the
high-ionisation gas in this object; (iii) the modelling of the observed emission gas spectrum at different locations of the galaxy taking into account the combine effects of both the radiation from the central source and shocks produced in the ISM by the passage of the radio jet. Notice that different phases of the outflow already detected in IC\,5063 have already been published \citep{tadhunter_2014, morganti_2015, dasyra_2016, venturi_2021} but, to the best of our knowledge, none of them have focused on the one traced by the coronal emission, which in turn, can be related to the X-ray phase. Throughout the paper, $H_{\rm 0}$ = 70~km~s$^{-1}$~Mpc$^{-1}$, $\Omega_{\rm m}$ = 0.30, and $\Omega_{\rm vac}$ = 0.70, have been adopted. At the redshift of IC~5063, $z$~=~0.01135  \citep{veron_2006}, the projected scale is 1$\arcsec$ = 239~pc.
 
In Sec.~\ref{sec:obs} we describe the data used for IC\,5063. In Sect.~\ref{sec:extension}, we characterise the coronal gas extension, geometry and morphology of the coronal line region (CLR) as well as the relationship with emission at other wavelength bands. In Sect~\ref{sec:kinematics}, we describe the gas kinematics of the galaxy, while in Sect.~\ref{sec:physical_cond} we study the physical conditions of the coronal gas. In Sect.~\ref{sec:models} we model the observed line ratios measured in this object. Final remarks are in Sect~\ref{sec:final}.  

\section{Observations}
\label{sec:obs}

VLT/MUSE data for IC\,5063 were downloaded from the MUSE Science Archive. To this purpose, the ESO file ADP.2016-07-22T15:57:54.788 (ESO programme observation ID: 60.A-9339(A)) observed on the night of June 23, 2014 was employed. The total exposure time was 2240~s, with an average seeing of 0.782$\arcsec$.   The IFU cube covers 1$\arcmin \times 1\arcmin$ with a sampling of 0.2\arcsec/spaxel. It is restricted to the wavelength range 4750–9350~\AA.  Here, we adopted the reduced cube provided by the ESO Quality Control Group. This reduction follows the standard steps of bias subtraction, flat-field correction and wavelength calibration. A standard star was also observed, and was used for flux calibration and telluric atmospheric correction.

The data analysis were carried out using custom {\sc python} scripts, following an approach similar to that described in \citet{ardila_2020} and \citet{fonseca-faria_2021}. Below, we briefly describe the procedures employed.
  
First, we fitted and subtracted the stellar continuum in the range 4770-7350 \AA. We did not extend that procedure to the red edge of the spectra because of the strong sky residuals redwards of 7400~\AA, which considerably affects the stellar synthesis results. The fitting of the stellar population procedure is necessary to fully recover the fluxes 
of the H\,{\sc i} lines affected by the underlying stellar population. Moreover, some weak lines may also be strongly 
diluted by the stellar continuum. To this purpose, the stellar population synthesis code {\sc starlight} \citep{cid_2005} 
was employed, together with the set of stellar populations of E-MILES \citep{vaz_2016}. Afterwards, the extinction correction due to  the Galaxy \citep[A$_{\rm V}$=0.19,][]{schlafly_2011} was applied. In this process, the CCM extinction law \citep{cardelli_1989} was employed.

In order to measure the flux, centroid position and the full width at half maximum (FWHM) of the emission lines at each 
spaxel we fitted Gaussian functions to individual lines or to sets of blended lines. This procedure was carried out using a set of custom 
scripts written in {\sc python} by our team.  
One or two Gaussian components were necessary to reproduce the observed profiles. 
 The criterion for the best solution among the one- and two-component ones was the minimum value of the reduced ${\chi}^{2}$. When a single line required more than one component, the one with the lowest FWHM was called "narrow" component while that with the largest value, was named "broad" component.  Moreover, some 
constraints were applied. 
For instance, doublets such as [\ion{N}{ii}]\,$\lambda\lambda6583,6548$ and [\ion{O}{iii}]\,$\lambda\lambda$4959,5007 were constrained to their theoretical line flux ratio. Also, lines belonging to the same doublet were constrained to have the same width and intrinsic wavelength 
separation. The [\ion{S}{ii}]\,$\lambda\lambda$6716,6731  doublet was constrained to have the same width and a relative theoretical wavelength separation of 14.4~\AA. Using that approach, the strongest emission lines detected in the data cube  (that is, H$\beta$, [\ion{O}{iii}]\,$\,\lambda\lambda$4959,5007, [\ion{Fe}{vii}]\,$\,\lambda$6087, [\ion{O}{i}]\,$\,\lambda\lambda$6300,6364, [\ion{Fe}{x}]\,$\lambda$6375,  H$\alpha$, [\ion{N}{ii}]\,$\,\lambda\lambda$6548,6584, and [\ion{S}{ii}]\,$\lambda\lambda$6716,6731) were fitted from the continuum-subtracted cube. We followed a similar approach to measure weak lines such as [\ion{S}{iii}]$\,\lambda6312$,  [\ion{Fe}{xi}]~$\lambda$7892,  and [\ion{O}{ii}]~$\lambda$7320. We note that [\ion{Fe}{xi}]~$\lambda$7892 and [\ion{S}{iii}]~$\lambda$9069 are located redwards of 7350~\AA. In these two cases, we interpolate the adjacent continuum by a spline curve of order 3 in order to measure the fluxes of both lines. This procedure is justified as they are not close in wavelength to any \ion{H}{i} or strong stellar absorption features. 

In order to determine the extinction, quantified by the E(B-V), we use Equation~1 in \citet{fonseca-faria_2021}, which takes into account the CCM extinction law \citep{cardelli_1989} and the observed and intrinsic H$\alpha$/H$\beta$ emission line flux ratio. We measured the observed H$\alpha$ and H$\beta$ fluxes at the spaxels where both emission lines are detected at 3$\sigma$-level. The intrinsic value adopted was 3.1 assuming case B recombination. It is more suitable to the typical gas densities found in AGN \citep{osterbrock_2006} and accounts for the effect of collisional excitation.  
At the spaxels where a narrow and a broad component were fit, for consistency we used  only the fluxes of the narrow components.  We found our extinction map very consistent with the one presented by \citet{mingozzi_2019} for IC\,5063 and for that reason it will not be shown here.

\begin{figure*} 
    \centering
    \includegraphics[width =17cm]{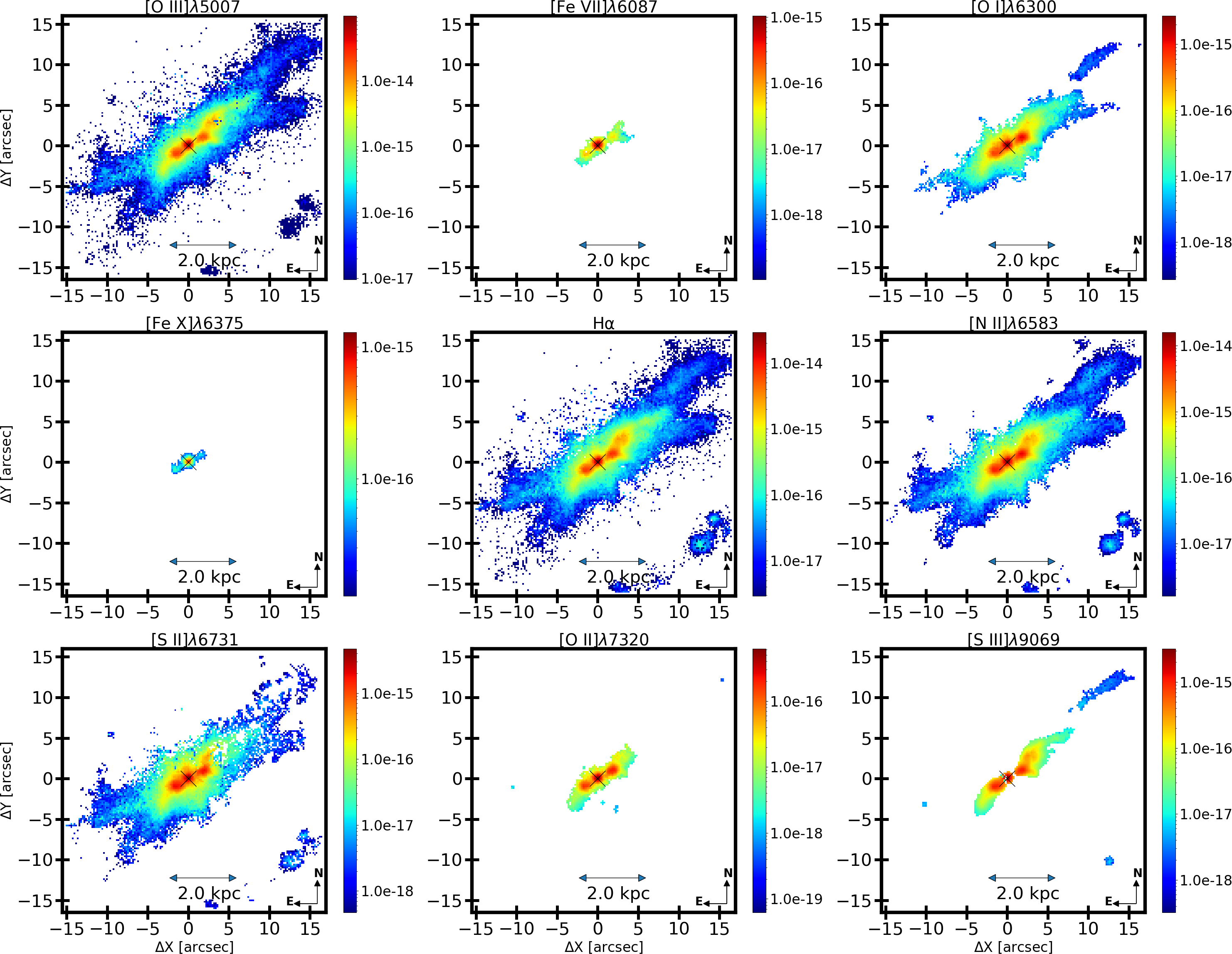}
    \caption{Extinction corrected emission-line flux distributions of IC\,5063. The colour bars (right side of each panel) show the fluxes in logarithmic scale and in units of erg\,s$^{-1}$\,cm$^{-2}$\,spaxel$^{-1}$. The white regions correspond to locations where the lines where not detected or the S/N $<3\sigma$.}
    \label{fig:ic_fluxmaps}
\end{figure*}

\begin{figure}
    \includegraphics[width=9cm]{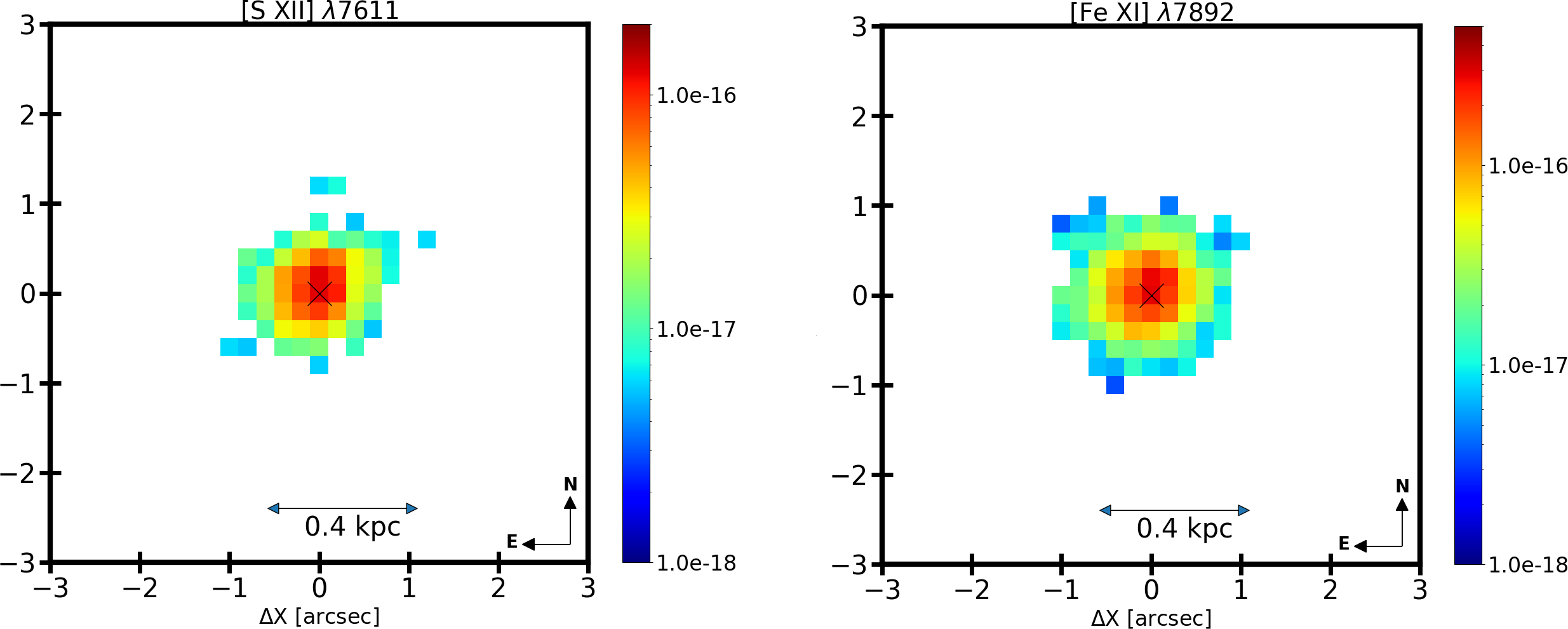}%[width=8.2\textwidth]
    \caption{   Flux map of the coronal lines [\ion{S}{xii}]~$\lambda$7611  (left panel) and [\ion{Fe}{xi}]~$\lambda$7892 (right panel). The colour bar show the fluxes in logarithmic scale and in units of erg\,s$^{-1}$\,cm$^{-2}$\,spaxel$^{-1}$.  In both cases, the emission region is spatially unresolved.}
    \label{fig:coronallines}
\end{figure}

Figure~\ref{fig:ic_fluxmaps} displays the flux distribution of the most relevant lines to this work in IC\,5063, obtained after the Gaussian fitting procedure. In all cases, the flux values of the different lines measured in each spaxel were corrected by E(B-V) as described above.  
It can be seen from Figure~\ref{fig:ic_fluxmaps} that overall, the gas emission extends along the SE-NW direction. Our maps agree very well to those in common shown in \citet{venturi_2021}.   The maps corresponding to [\ion{Fe}{vii}]~$\lambda$6087  (first row, second column), [\ion{Fe}{x}]~$\lambda$6375  (second row, first column), and [\ion{O}{ii}]~$\lambda$7320 (third row, second column), to the best of our knowledge, were not published before. In addition, coronal emission belonging to [\ion{S}{vii}]~$\lambda$7611 and [\ion{Fe}{xi}]~$\lambda$7892  was detected only at the nucleus and it was not angularly resolved. Figure~\ref{fig:coronallines} displays the corresponding flux map distribution.

In this work, we focus on the coronal emission of IC~5063, the gas kinematics, degree of ionisation and shock effects using a  large set of emission lines, including high ionisation lines.

\section{The coronal line region in IC~5063}
\label{sec:extension}

The study of the coronal gas extension in AGN is a key piece to register the most energetic component of the ionised gas that participates in the process of \outs. It also allows us to investigate the driving mechanism in the gas ionisation (photoionisation by AGN or by shocks or both). To this purpose, we selected the gas emission traced by the lines of [Fe\,{\sc vii}]\,$\lambda6087$ and [Fe\,{\sc x}]\,$\lambda6375$ (hereafter [Fe\,{\sc vii}] and [Fe\,{\sc x}]) as representative of the high-ionisation gas. The presence of both lines in the nuclear spectrum of IC\,5063 was first reported by \citet{Colina91} by means of long-slit spectroscopy. They also reported extended [Fe\,{\sc vii}] emission within the inner 1$\arcsec$.5 around the AGN. [Fe\,{\sc vii}] and [Fe\,{\sc x}] have an ionisation potential of 100~eV and 274~eV, respectively. The high energy required for their production implies that they cannot be produced by stellar processes, as is the case of [\ion{O}{iii}]~$\lambda$5007, which may have a contribution from star-forming regions.

\begin{figure*} 
    \centering
    \includegraphics[width =18.2cm]{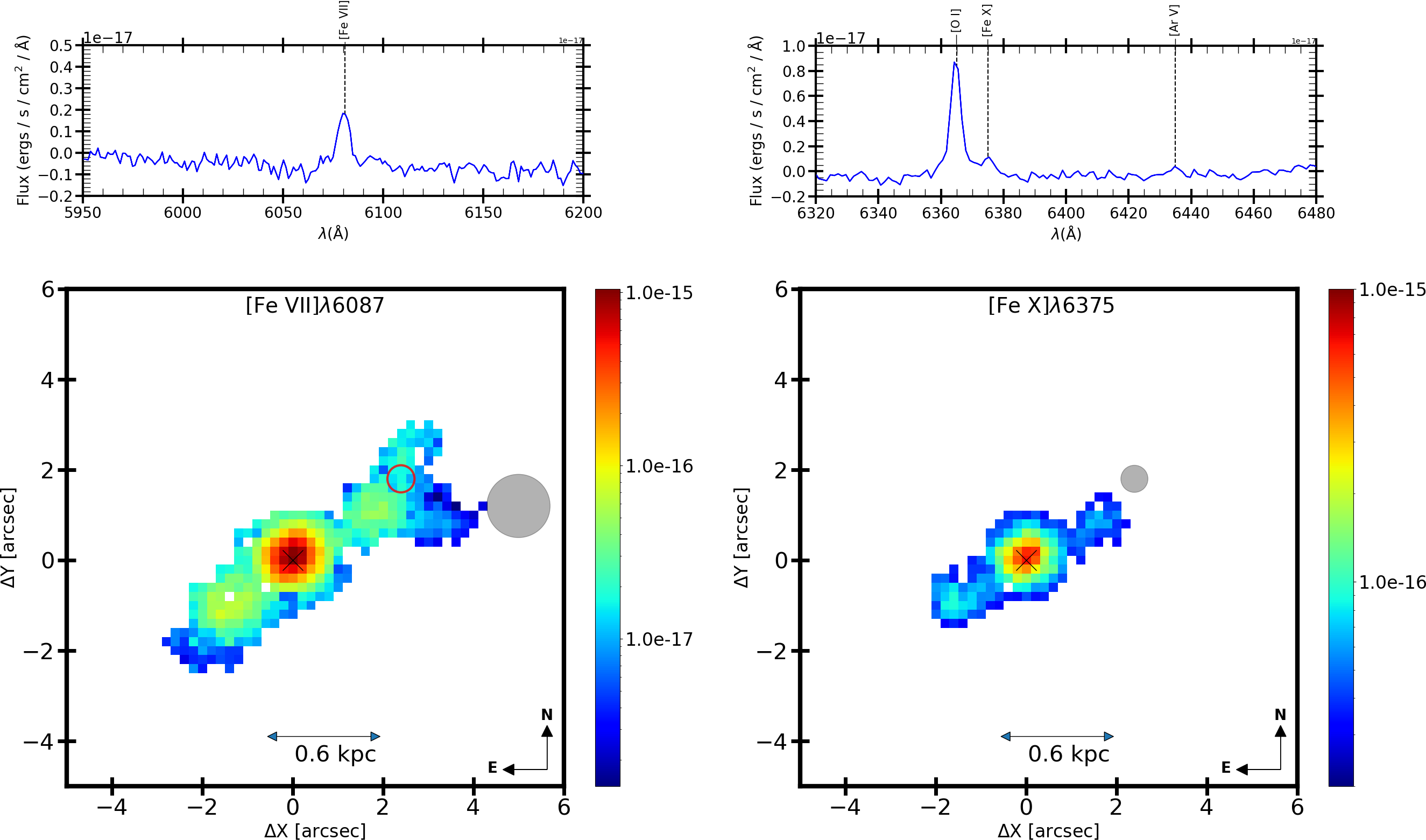}
    \caption{ Bottom: Zoom-in of the coronal emission flux distribution shown in Fig.~\ref{fig:ic_fluxmaps} for  [\ion{Fe}{vii}] (left) and [\ion{Fe}{x}] (right). The upper two panels show the corresponding emission line profile at the largest distance from the AGN where they were detected grey circles in the bottom panels; their sizes show the region integrated to obtain the observed emission line profile. The red open circle in the [\ion{Fe}{vii}] map shows the region of maximum extension of [\ion{Fe}{x}].  The white regions correspond to locations where the lines where not detected or the S/N $<3\sigma$. The colour bars at the right of each panel) show the fluxes in logarithmic scale and in units of erg\,s$^{-1}$\,cm$^{-2}$\,spaxel$^{-1}$}
    \label{fig:extIC5063}
\end{figure*}

Figure~\ref{fig:extIC5063} shows the coronal emission found in IC\,5063. In the two panels to the left, the emission flux distribution of [\ion{Fe}{vii}] (bottom) and the spectrum collected at the largest distance from the AGN  are presented. It can be seen that at the AGN position (marked with the black cross) that emission reaches its maximum intensity, amounting to $10^{-15}$~erg\,s$^{-1}$\,cm$^{-2}$\,spaxel$^{-1}$. In addition, extended coronal emission is clearly detected. It is distributed in an elongated structure along the northwest-southeast direction, with a total extension of $\sim$2~kpc from both opposite ends. 

In the extended region, two secondary peaks of coronal emission are identified at 500\,pc to the northwest and at 400\,pc to the southeast of the AGN. Both peaks have a maximum flux of $\sim10^{-16}$~erg\,s$^{-1}$\,cm$^{-2}$\,spaxel$^{-1}$. In the secondary peak to the NW of the AGN, the emission is clearly splitted, following a Y-pattern that resembles a fork. The base of the Y-feature points towards the AGN and the secondary peak of emission coincides with the bifurcation point of the fork, from where two arms are observed,  one to the SW and another to the NW. The arm to the SW is the most extended one. Notice that the Y-feature coincides in position and morphology to the one already identified by other authors using low- to mid-ionisation lines \citep{morganti_2007, dasyra_2015,venturi_2021}. 

The most distant [\ion{Fe}{vii}] emission is identified at a distance of 1193$\pm$39~pc from the AGN, in the tip of the SW arm of the Y-pattern. This result is found after summing up the signal in a circular aperture of  0.6\arcsec in radius. A similar procedure was carried out in the NW arm of the fork but no emission was detected beyond what is observed in Figure~\ref{fig:extIC5063}. We do not discard that the 
 NW branch of the fork could be more extended than we see. However, the combined effect of dust extinction and low S/N in this region hinders its detection.  

To the SE of the AGN  [\ion{Fe}{vii}] emission is detected up to $\sim$\,680~pc. No signs of split emission is observed. Outwards of that secondary peak, the emission bends slightly to the South and fades out. Outside the nucleus and the two adjacent peaks, the extended emission has an average flux of $\sim10^{-17}$~erg \,s$^{-1}$\,cm$^{-2}$\,spaxel$^{-1}$. 

The two panels to the right in Figure~\ref{fig:extIC5063} present the emission map for [\ion{Fe}{x}] (bottom) and the spectrum at the largest distance from the AGN (top). This emission peaks at the nucleus, with secondary peaks in the circumnuclear region, around the two off-nuclear radio spots. We therefore conclude that the [\ion{Fe}{x}] gas is spatially resolved although  it is more compact than that of [\ion{Fe}{vii}]. In the region coinciding with the AGN, it displays a peak flux of $1\times10^{-16}$~erg\,s$^{-1}$\,cm$^{-2}$\,spaxel$^{-1}$. Outside the nuclear component, we noticed enhanced emission of [\ion{Fe}{x}] that matches in position the secondary peaks detected in other lines. The maximum extension of the [\ion{Fe}{x}] gas in the unbinned cube is detected at $\sim$450~pc NW of the AGN. However, we scanned the entire MUSE cube and identified [\ion{Fe}{x}] emission at a maximum distance of 696$\pm$46~pc NW from the AGN, after integrating the signal in a circular aperture of 0.2\arcsec in radius.  It coincides with the start of the NW arm in the Y-shape feature detected in the [\ion{Fe}{vii}] line. To the best of our knowledge, it is the first time in the literature that extended coronal emission in the [\ion{Fe}{vii}] and [\ion{Fe}{x}] lines is reported in IC\,5063. Previously, \citet{dasyra_2015} found extended high-ionisation emission of [\ion{Si}{vi}]~1.963\,$\mu$m in the $K-$band. The extension and morphology of that emission is very similar to that observed in [\ion{Fe}{vii}] by us. It is important to highlight that the extended coronal emission is also co-spatial with the emission of [\ion{O}{iii}], although the latter is significantly more extended (see Figure \ref{fig:ic_fluxmaps}).

As mention previously, we report the detection of unresolved coronal emission of  [\ion{S}{vii}]~$\lambda$7611 and [\ion{Fe}{xi}]~$\lambda$7892. The ionisation potential of these lines are 281~eV and 233~eV, respectively. In both cases, the emission region is restricted to the spatial resolution of the observation (0.8$\arcsec$). Their integrated line profiles  lack the complexity  seen in [\ion{Fe}{vii}] or [\ion{Fe}{x}] and their total flux is $<$1\% of [\ion{O}{iii}].

\subsection{The CLR and its relationship with the radio jet}
\label{sec:clrandjet}

\begin{figure}
    \includegraphics[width=8cm]{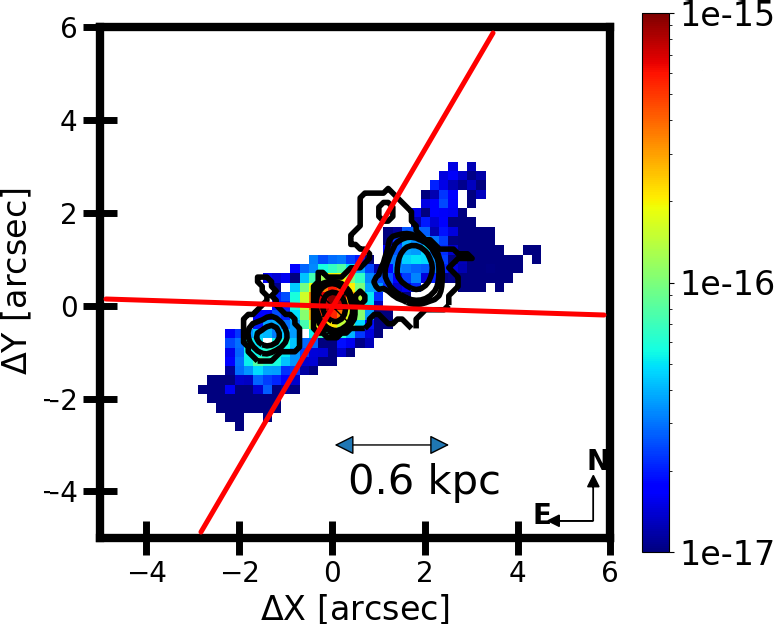}%[width=8.2\textwidth]
    \caption{
    [\ion{Fe}{vii}] flux map with radio contours overlaid. In red we plot the edges of the ionisation cone as defined by \citet{travascio_2021}.  The colour bar shows the emission line flux in logarithmic scale and in units of erg\,s$^{-1}$\,cm$^{-2}$\,spaxel$^{-1}$. 
    }
    \label{fig:gals_radio}
\end{figure}

In this section, we analyse the radio emission in IC\,5063 and its relation with the coronal gas extension. To this purpose, we employ observations taken by the ATCA (Australia Telescope Compact Array) radio telescope at 17.8 GHz. Information about these observations and data reduction are in  \citet{morganti_2007}.   
Figure~\ref{fig:gals_radio},  shows the radio contours (in black) overlaid to the map of the [Fe\,{\sc vii}] flux distribution.

It is possible to see  that the extended radio emission in IC\,5063 at 17.8~GHz is more compact than that of [\ion{Fe}{vii}] (and all other emission lines). However, notice that the radio lobes to the NW and SE of the AGN (as well as at the AGN itself) is co-spatial or very close to the most intense spaxels detected in [\ion{Fe}{vii}].  Indeed, the radio emission at the two lobes is surrounded by intense coronal emission, as can be observed in Figure \ref{fig:gals_radio}.

\begin{figure*}
    \includegraphics[width=8cm]{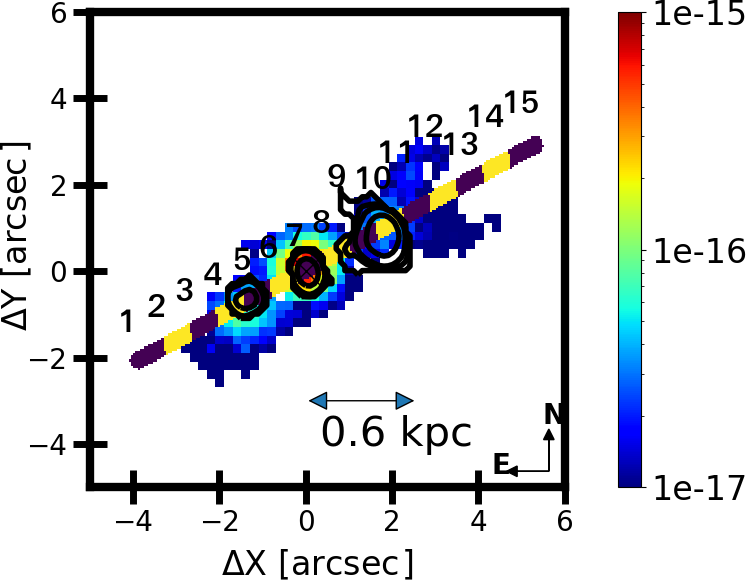}%[width=8.0\textwidth] 
    \includegraphics[width=7.5cm]{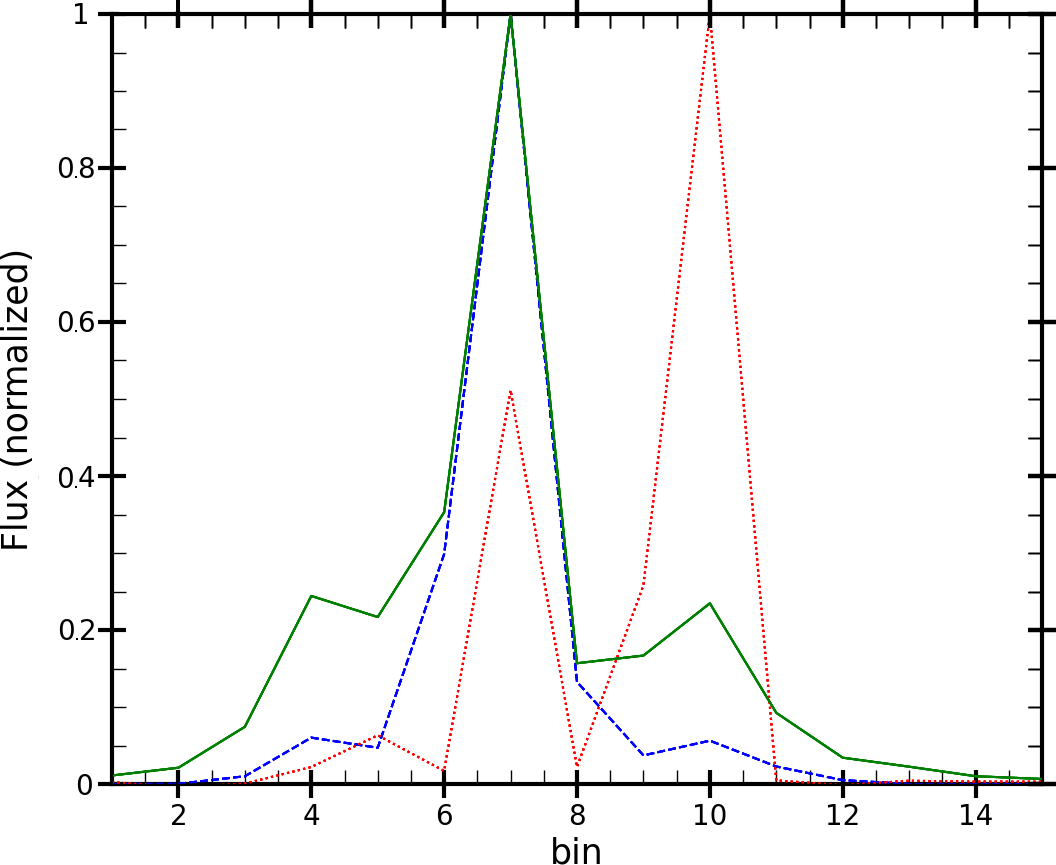}%[width=7.5\textwidth]
    \caption{Left panel: [\ion{Fe}{vii}] flux map with radio contours overlaid. The pseudo-slit with segments in purple and yellow marks extraction windows employed to map the radio and line emission along the spatial direction. The aperture number of each bin is indicated in black.
    The colour bar shows the emission line flux in logarithmic scale and in units of erg\,s$^{-1}$\,cm$^{-2}$\,spaxel$^{-1}$. A similar procedure was done in the [\ion{O}{iii}] map.
    Right panel: Normalised flux of the radio emission (red dotted line), the [\ion{Fe}{vii}] emission (dashed blue line), and the [\ion{O}{iii}] emission (full green line)  measured along the apertures shown in the left panel. The x-axis is numbered following the aperture numbers indicated in the left panel.}
    \label{fig:gals_radiofe7_bins}
\end{figure*}

In the left panel of Figure \ref{fig:gals_radiofe7_bins}, a pseudo-slit mask is overlaid to the [\ion{Fe}{vii}] map and the radio emission (black contours). The position angle of the slit coincides to that of the radio axis. Integrated values of the flux were extracted along each of the 15 bins in both maps, with the first one at the SE edge of the slit and the last one at the NW edge. A similar procedure was applied to the [\ion{O}{iii}] map. We aimed at comparing the flux distribution of the radio jet and the mid- and high-ionised gas. For instance, we want to confirm if they all peak at similar spatial positions along the radio axis.

The right panel of Figure~\ref{fig:gals_radiofe7_bins} shows the flux distribution measured in each bin for the radio continuum (dotted red line), [\ion{Fe}{vii}] (dashed blue line) and [\ion{O}{iii}] (full green line) emission. Maximum flux values have been normalised to 1. It can be seen that the peaks of the line emission and radio continuum at bins 7 and 10 (AGN and the northwest radio lobe, respectively) coincides spatially. The strongest radio emission peak is observed at the NW radio lobe. As shown by \citet{dasyra_2016}, this region displays a stronger interaction with the radio jet. Previously, \citet[][and references therein]{morganti_2015} identified a strong outflow at the same position. To the southwest, both emission lines peak at bin 4 while the corresponding radio lobe is observed at bin 5. Still, the SE radio continuum peak is the faintest of the three. Figure~\ref{fig:gals_radiofe7_bins} reveals the strong association between the high-ionisation emission gas and the radio jet, both at the nucleus as well as in the extended emission.

In addition to the radio jet, extended soft and hard X-ray $Chandra$ emission (0.3-7.0 keV) is detected in IC\,5063 \citep{travascio_2021}. By a simple visual inspection of Figures~2 and~3 in \citet{travascio_2021}, the X-ray emission is aligned and co-spatial with the coronal gas. These authors modelled the X-ray emission and found that it is consistent with a phase of low ionisation (log $U \sim$ -1.7, $-$ 2.7) and less
obscured ($N_{\rm H} < 10^{22}$~cm$^{-2}$) gas with respect to the nuclear component, plus a more ionised (log $U \sim$ 1.8)
phase of the gas or a collisionally excited gas with
kT $\sim$ 1 $-$ 1.3 keV. They conclude that the increase of soft X-ray emission along the jet suggests jet-ISM interaction
as a likely trigger for most of the  X-ray emitting gas.

\section{Coronal gas kinematics}
\label{sec:kinematics}

\begin{figure*}
\centering 
    \includegraphics[width=16.0cm]{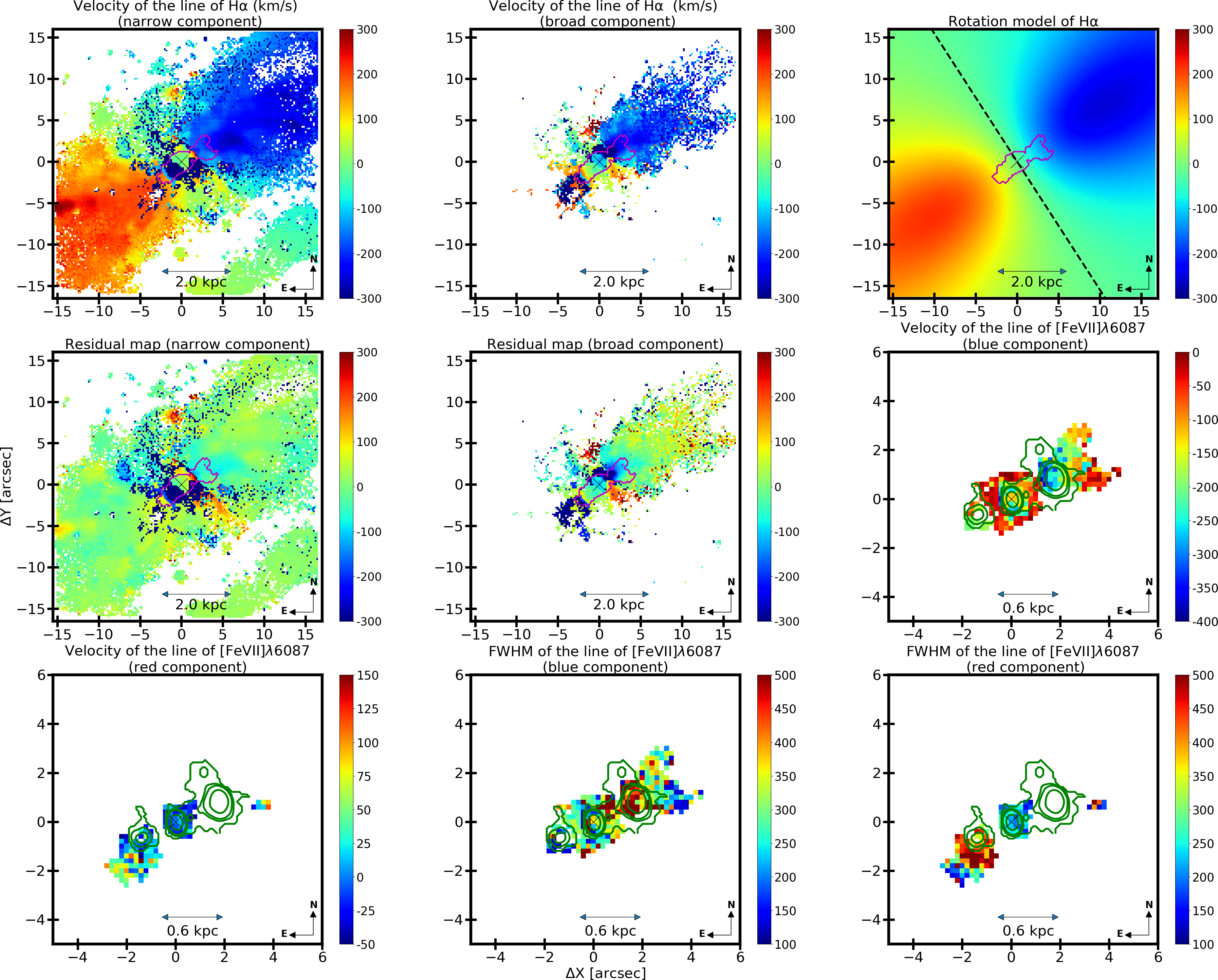}
    \caption[IC gas kinematics maps\,5063]{Kinematics maps for IC\,5063. The first row presents, respectively, the velocity map for the narrow component of H$\alpha$ (left), the velocity map for the broad component of that line (centre) and the rotation model fit for the narrow component of H$\alpha$ (right). The dashed line indicates the position angle of the line of nodes ($\Psi_{0}$) obtained from the fit. In the second row we show the residual map of the narrow and broad components of H$\alpha$ (left and centre panels, respectively), obtained after subtraction of the best fit rotation model from the observed H$\alpha$ velocity. The contour in magenta marks the [\ion{Fe}{vii}] emission region and the contour in green marks the radio emission region. The velocity map of [\ion{Fe}{vii}] (blue component) is in the right panel. In the third row we present the velocity map of [\ion{Fe}{vii}] (red component), and the FWHM distribution of [\ion{Fe}{vii}] blue component (centre) and red component (right).}
    \label{fig:cinematica_ic5063}
\end{figure*}

From the emission line fitting described in Sect.~\ref{sec:obs}, we constructed velocity maps for the H$\alpha$ and [\ion{Fe}{vii}] emission detected in IC~5063.  Our choice of H$\alpha$ is because the narrow component of this line usually describes best the galaxy disk rotation, that we want to model here.
The first row of Figure \ref{fig:cinematica_ic5063} presents the results for the narrow (left panel) and broad (center panel) component of the former line. The contour line in black marks the region containing the bulk of the  [\ion{Fe}{vii}] emission. 

 We fit to the observed line of sight velocity of the H$\alpha$ line the rotation model  described in  \citet{bertola_1991}, which considers that the gas is on circular orbits 
 in a plane. Only the narrowest component of that line is used in the fit. This is because that component is observed in most of the MUSE field and, in general, it is more related to disc emission.  
 More details about the fitted model can be  seen in Equation 4 in \citet{fonseca-faria_2021}.

The  modelled rotation field for the gas kinematics associated to the narrow component of H$\alpha$ is in the first row, third column of Figure~\ref{fig:cinematica_ic5063}. In the panels at the second row, first and second columns, we present the residual maps (observed velocity  minus the rotation model) for the narrow and broad component of H$\alpha$, respectively.
The remaining four panels (second row, third column and the three panels in the third row) present, respectively, the residual maps of the blue and red components of the [\ion{Fe}{vii}] line and their corresponding FWHM.   By red and blue components we indicate the position of the peak of the line profile. When the peak is to the red (blue) of the systemic velocity, we identify it as the red (blue) component. Notice that the residual maps of [\ion{Fe}{vii}] were constructed after subtraction from the observed velocity, at every spaxel, the best rotation model of the narrow component of H$\alpha$.

From the velocity map of the narrow component of H$\alpha$ shown in Figure \ref{fig:cinematica_ic5063}, it can be seen that in a circular region centred on the nucleus and with a radius of 5$\arcsec$, the gas velocity is highly complex and does not follow the disc rotation. Indeed, velocities in excess of $\pm$200~\kms\ are observed. Outside that central region, the rotation model  satisfactorily reproduces the velocity field, with velocity residuals of $\pm$30~\kms\ at the most. The residual velocity map for the broad component of H$\alpha$ shows a similar behaviour, with values in excess of 200\,\kms\   compared to the rotation model. It is important to emphasise that the regions where the H$\alpha$ emitting gas is more strongly disturbed coincide spatially with the NW and SE radio lobes. This behaviour supports the hypothesis of an interaction between the jet and the interstellar medium and confirms recent results gathered by \citet{mukherjee_2018}. They modelled this interaction in IC\,5063 and found that the jet ablates, accelerates, and disperse clouds to velocities exceeding 400~\kms. Their results strongly supports the hypothesis that the jet is responsible for the strongly perturbed gas dynamics, seen here in the ionised component.   

Similar results are obtained here in the highest ionised component. The velocity maps of [\ion{Fe}{vii}] show that this line displays double-peak profiles mainly in the nuclear region. From Figure \ref{fig:cinematica_ic5063}, it is easy to see that the coronal gas is decoupled from the rotation of the disc of ionised gas. In fact, velocities in excess of up to  400\,\kms\  in relation to the gas rotation are identified and coincident with the position of the radio lobes. At these locations, high FWHM values are identified, of up to 500\,\kms, which coincide with high flux values of [\ion{Fe}{vii}] (see Figure \ref{fig:cinematica_ic5063}). These results may indicate the presence of an expanding and highly turbulent gas. 
At this point, it is also important to mention the large velocity dispersion observed in the ionised gas perpendicular to the jet \citep{dasyra_2015, venturi_2021}, which seem to peak close to the NW radio lobe where the FWHM of [\ion{Fe}{vii}] has its maximum values, though this gas does not extend perpendicular to the jet. They interpret these lateral outflows as due to the action of the jet perturbing the gas in the galaxy disk. This result suggests that low-power jets are indeed capable of affecting the host galaxy.

We also construct channel maps, which allow us to examine the velocity field of a spatially distributed gas. This technique consists of slicing a given emission line into bins of velocity, so that each bin or channel map shows the regions that simultaneously present a given velocity. Since our objective is to examine the behaviour of the extended high-ionisation gas, we will use the line of [Fe\,{\sc vii}]\,$\lambda6087$ as representative of this gas. In this process, we assume a \textit{bin} between two consecutive channels corresponding to $\Delta \lambda$\,=\,1.25\,\AA\ or 61.5\,\kms\ in velocity space. In the channel maps, line fluxes are measured if they have at least twice as much flux as the standard deviation in the adjacent residual continuum of the line. The adopted continuum region has a width of $\sim$\,$30$\,\AA\ to the red and blue of the emission line of [\ion{Fe}{vii}].    

\begin{figure*}
    \centering 
    \includegraphics[width=17.0cm]{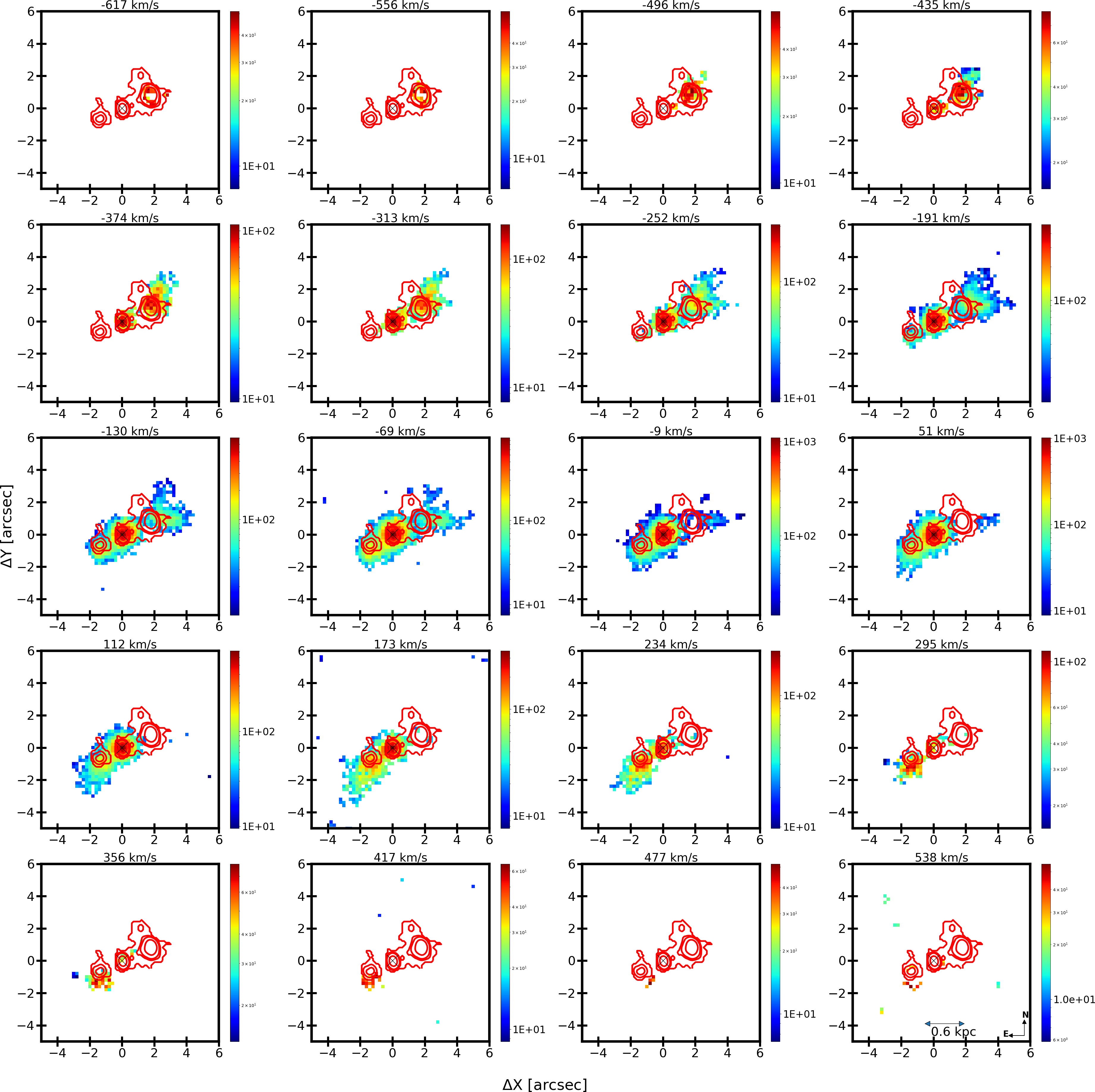} 
    \caption{Channel maps of the line of [Fe\,{\sc vii}]\,$\lambda6087$. The velocity shown at the top of each panel represent the central velocity within a bin of 61.5~\kms\ of width. Fluxes are presented in units of $10^{-20}$\,erg\,s$^{-1}$\,cm$^{-2}$\,spaxel$^{-1}$.  The contours in red mark the radio jet.}
    \label{fig:channels_ic5063}
\end{figure*}

In the channel maps of the [\ion{Fe}{vii}] emission (see Figure~\ref{fig:channels_ic5063}),  the
gas clearly departs from the galaxy disc rotation. At the AGN position, we see a broad blue-asymmetric profile,  which appears as an emission feature from channel $-$435\,\kms\ to 356\,\kms.  This result is consistent with the velocity maps of [\ion{Fe}{vii}] presented in Figure \ref{fig:cinematica_ic5063}, where the blue component has more   extreme velocity (-400 \kms) when compared to the red component, which has a maximum speed of 150\kms.  It also shows that the NW lobe is dominated by gas with negative velocities, being observed between the channels $-$617\,\kms\ to 51\,\kms. It implies that the bulk of this gas is approaching the observer. Notice also that in the slices corresponding to the velocities in the interval -191~\kms\, to 51~\kms, the region showing the most distant [\ion{Fe}{vii}] emission is clearly observed.  

The coronal emission in the SE lobe is formed by gas that  mostly moves away from the observer, being identified between the channels $-$191\,\kms\ and 538\,\kms. In this region, we see splitted line profiles, particularly in and around the radio lobe. The slight bending to the south of the gas emission that is  farther out of the radio lobe is apparent in the slices 112~\kms, 173~\kms, and 234~\kms. Excess velocities of 350~\kms\ relative the systemic are found. Therefore, although this high-ionisation gas is very close to the galaxy disc, the evidence gathered from the co-spatiality between radio lobes, optical line lobes and coronal line lobes as well as the results from the simulations from \citet{Mukherjee_2018b} tailored at IC\,5063  suggest that it is being accelerated by the presence of the radio-jet.

The analysis of Figure \ref{fig:channels_ic5063} made above, shows as main result that high-velocity high-ionisation gas near the projected positions of both radio lobes are present in IC\,5063. From our results and others reported in the literature \citep{dasyra_2015, venturi_2021}, very likely this high-ionisation gas is accelerated by the passage of the radio jet. Notice that in low-ionisation lines, we as well as the authors above, observed gas accelerated in the direction perpendicular to the radio jet axis, mainly in the cross-cone region (see Figure~\ref{fig:cinematica_ic5063}, leftmost and middle panels of second row). This result reinforces the role of the jet in shaping the gas kinematics in the central region and along the jet direction in this AGN.

\section{Physical Conditions of the highly ionised gas}
\label{sec:physical_cond}
 
\subsection{Ionisation degree of the gas}
 \label{subsec:degree_gas}

Studies of the degree of ionisation of a gas cloud aim to determine, among others,  the nature of the photon source that illuminates it. To this purpose, line ratios between ions are used, usually of the same element or relative to hydrogen, to avoid potential biases introduced by metallicity or differences in critical density of the transitions involved. With this in mind, we use in this work the lines of [N\,{\sc ii}]\,$\lambda6583$, [O\,{\sc iii}]\,$\lambda5007$ and [Fe\,{\sc vii}]\,$\lambda6087$ as representatives of low, medium, and high ionisation gas, respectively. 

\begin{figure*}
    \includegraphics[width=17.0cm]{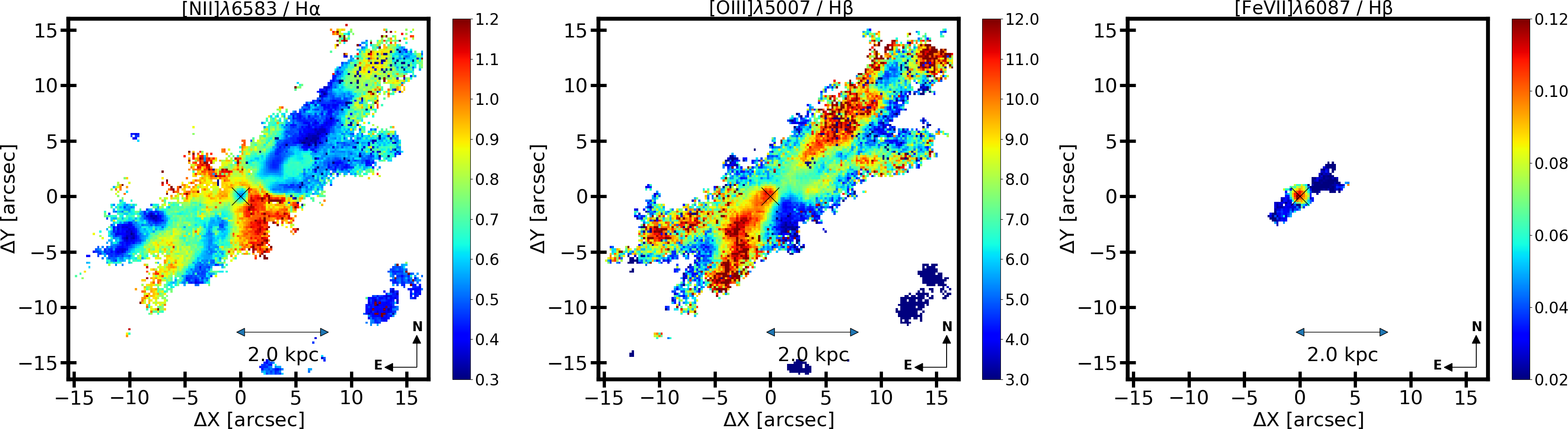} 
    \centering
    \caption[  Emission line flux ratio maps for IC\,5063.]{Emission line flux ratio maps [N\,{\sc ii}]\,$\lambda6583$\,/\,H$\alpha$ (left), [O\,{\sc iii}]\,$ \lambda5007$\,/\,H$\beta$ (centre) and [Fe\,{\sc vii}]\,$\lambda6087$\,/\,H$\beta$ (right).}
    \label{fig:razaolinhasgalaxias}
\end{figure*}

In Figure \ref{fig:razaolinhasgalaxias}, the emission line flux ratio of the former line relative to H$\alpha$ and the latter two relative to  H$\beta$ are presented. We aim at comparing the extension and variation of the values of each ratio within the region where high-ionisation gas is detected.  Previously, \citet{sharp_2010, mingozzi_2019} and \citep{venturi_2021} had made a similar analysis but without the inclusion of the coronal iron line.

High values of [O\,{\sc iii}]\,/\,H$\beta$ ($>$\,6) are observed in the region that is co-spatial to [\ion{Fe}{vii}], that is, within the ionisation cone. Indeed, values as high as 12 are detected at $\sim$1~kpc from the AGN, where [\ion{Fe}{vii}] is no longer visible, suggesting that the ionisation state of the gas is kept nearly constant. In a scenario dominated by photoionisation by a central source, it is expected that the ratio [O\,{\sc iii}]/H$\beta$ decreases outwards from the AGN \citep{ferguson1997}.  In the direction of the NW radio lobe and across the Y-shape region, the above ratio displays values between 6 and 7. However, outside the [\ion{Fe}{vii}] emission region and following the two branches of the Y-shape, [O\,{\sc iii}]/H$\beta$ suddenly jumps to values of 8 and larger, following a stream that runs to the NW, parallel to the radio jet axis. This stream is surrounded by gas with ratios of $\sim$7.  In the cross-cone region (defined here as the region perpendicular to the major axis of the ionisation cone), [O\,{\sc iii}] is also detected. Its strength relative to H$\beta$ amounts to $\sim$7 to the NE. To the SW and to the South, it drops to values of $\sim$4, reflecting the decrease of the ionisation state of the gas.

[N\,{\sc ii}]\,/\,H$\alpha$ displays an opposite result to [\ion{O}{iii}]/H$\beta$. It reaches the highest values in the cross-cone region, particularly in the SW and south, with values close to $\sim$1.2. This is about twice the ratio measured at the AGN position. Towards the NE cross-cone region, values between 0.6 and 0.9 are detected. The lowest values of the  [N\,{\sc ii}]\,/\,H$\alpha$ ratio are measured along the two streams that extend beyond the Y-feature. 
 
[Fe\,{\sc vii}]\,/\,H$\beta$ is the largest at the AGN position and within the innermost 1$\arcsec$ region, with values between 0.06 and 0.1. Outside this region, the value of the ratio decreases by less than one dex, remaining quite homogeneous, with values between 0.02 and 0.05, regardless of the distance to the nucleus.

In general, the line ratios studied here indicate that regions with a high degree of ionisation are associated to the ionisation cone. The comparison of Figure \ref{fig:razaolinhasgalaxias} with the radio maps presented in the previous section allows us to claim that these highly ionised regions are aligned with the radio jet. Maximum values of [O\,{\sc iii}]\,/\,H$\beta$ and [Fe\,{\sc vii}]\,/\,H$\beta$ are found at the AGN position and regions around it and along the radio jet. In contrast, [N\,{\sc ii}]\,/\,H$\alpha$ has the largest values in the regions outside the cone, perpendicular to the jet propagation direction. Low values of that ratio are identified where the [Fe\,{\sc vii}] is observed.

The emission of [Fe\,{\sc vii}], in addition to being more compact than that of [O\,{\sc iii}], is considerably more collimated than the latter. Because the coronal lines are not produced by stellar processes, the results show that this emission unambiguously traces the region of highest ionisation within the cone. As already shown in Sect.~\ref{sec:clrandjet}, the extended CLR follows, at least in projection, the jet emission. It is possible that the coronal gas extends farther out. However, deeper observations would be needed to confirm this hypothesis.

\subsection{BPT diagram}\label{sec:bpt}

We  analyse the gas excitation using the BPT diagnostic diagram log\,([O\,{\sc iii}]\,$\lambda5007$\,/\,H$\beta$) versus log\,([N\,{\sc ii}]\,/\,H$\alpha$) proposed by \citet{veilleux_1987}. The total flux associated to   these four lines was employed here. The diagram separates regions dominated by star formation (SF) from those dominated by photoionisation by AGN according to the values of the line ratios measured in different astronomical sources. In order to separate AGNs and star forming regions, we use the theoretical curve proposed by \citet{kewley_2001}.   The purpose of this plot is to spatially associate the spaxels where [\ion{Fe}{vii}] is detected to their position in the  diagram.

\begin{figure*}
\centering 
     \includegraphics[width=17.cm]{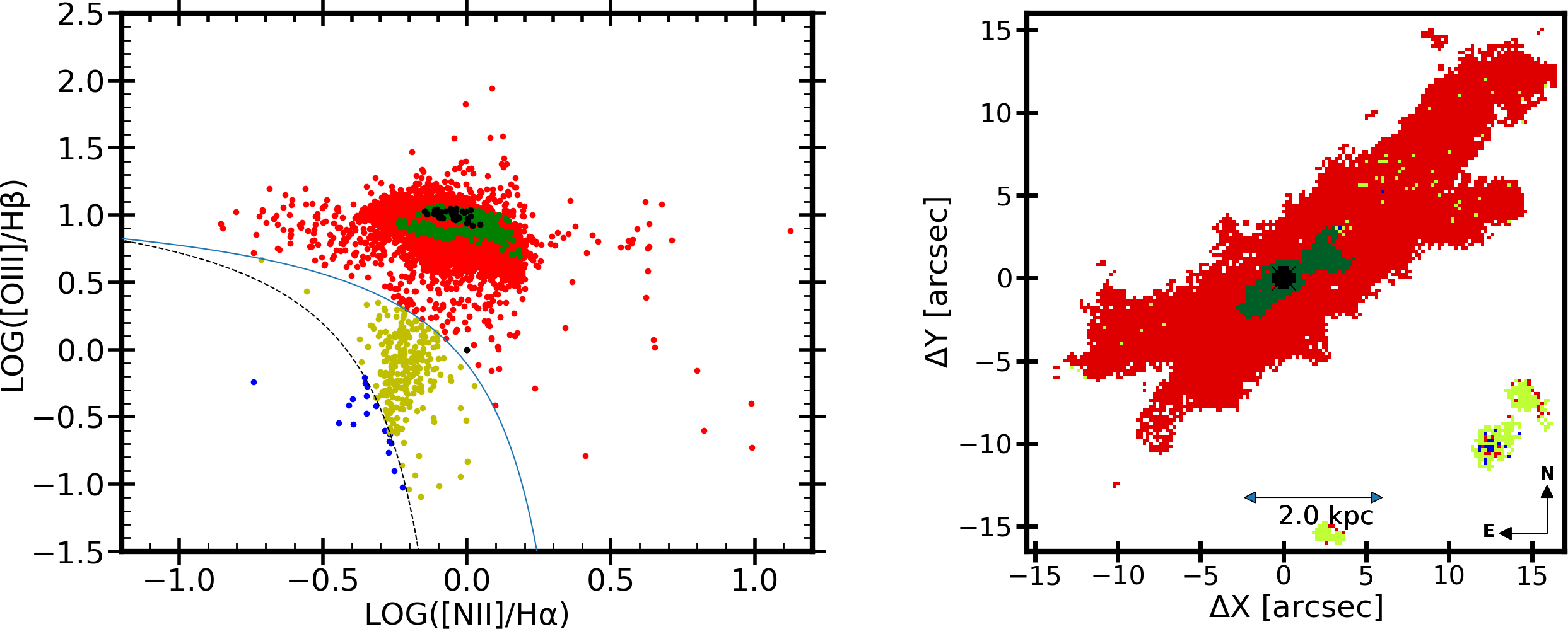}
     \caption[BPT diagram for IC\,5063.]{Left: BPT diagram log\,([O\,{\sc iii}]\,$\lambda5007$\,/\,H$\beta$) versus log\,([N\,{\sc ii}]\,$\lambda6584$\,/\,H$\alpha$). Right: spatial location in the galaxy of the spaxels plot in the left panel. The region dominated by the AGN is shown in red while the H\,{\sc ii} regions (star forming regions) are shown in blue. Spaxels with [Fe\,{\sc vii}] emission appear in green while the spaxels coinciding with the AGN PSF are in black. The black dashed curve is the boundary between star-forming galaxies and AGN as defined by \citet{kauffmann_2003}, while the solid blue line is the theoretical upper limit between starburst galaxies and AGN found by \citet{kewley_2001}.}
     \label{fig:f1bpt}
\end{figure*}

The left panel of Figure~\ref{fig:f1bpt} shows the BPT diagram for all the spaxels where the above four lines are 3$\sigma$ detected. The right panel shows the location of the spaxels employed to construct the diagram in the galaxy. This allows us to associate the dominant ionisation source with a spatial position in the object. The dots in blue represent spaxels with line ratios typical of SF regions while those in red represent spaxels where the ionisation source is associated with the AGN. In green we identify the spaxels  with [Fe\,{\sc vii}] emission and, in black, the spaxels associated with the nuclear region of the galaxy. The size of the latter was set as equal to the {\it seeing} measured during the observations.
 
We found that more than 95\% of the spaxels with [Fe\,{\sc vii}] emission occupy the locus of points dominated by AGN excitation. This result 
supports our claims that most of the ionised gas in the nuclear and circumnuclear region as well as along the jet propagation direction are compatible with gas ionisation by a hard continuum source  either from the AGN itself or by shock induced by the interaction of the radio jet with the ISM or both. We do not discard
the contribution of stellar processes, but if present, they are not  the dominant mechanisms in the gas excitation. Our results  confirm the ones obtained by \citet{sharp_2010, mingozzi_2019, venturi_2021} for this AGN. However, in this work we additionally mark the spaxels occupied by the coronal gas.

\subsection{Temperature of the ionised gas} \label{sec:gasden}

The wavelength range of the MUSE data allows us to determine the electron temperature using the lines of [\ion{S}{iii}]\,$\lambda$6312 and [\ion{S}{iii}]\,$\lambda$9069. We adopted a mean density of $n_{\rm e}$ = 5000~cm$^{-3}$ for the nucleus and the NW and SE lobes. That value was found by~\citet{HOLDEN_2022} using the [\ion{Ar}{iv}]~$\lambda\lambda$4711,4741 ratio for the nucleus and the NW lobe. These lines are more suitable to map the density of the mid-ionised gas,  when compared, for example, with that derived from the [SII] 6716/6731 ratio, as the ionisation potential necessary to produce the \ion{Ar}{iv} ion is 40.7~eV, close to the one required to   produce \ion{O}{iii} (35.2~eV).

Our temperature map is always limited to the region where the [\ion{S}{iii}]\,$\lambda$6312 line is observed. That line is considerably fainter ($\leq$ 10\%) than its counterpart at $\lambda$9069.  The [\ion{S}{iii}] lines are usually regarded as a suitable shocked gas diagnostics~\citep{osterbrock_2006}. It is also important to mention that the line of [\ion{S}{iii}]\,$\lambda$6312 is blended with the red wing of [\ion{O}{i}]\,$\lambda$6300 in most spaxels where the former line is observed. Although we were able to separate both lines,  the uncertainty of the results are large, mostly because of the weakness of [\ion{S}{iii}]\,$\lambda$6312.

\begin{figure}
\centering 
     \includegraphics[width=8.5cm]{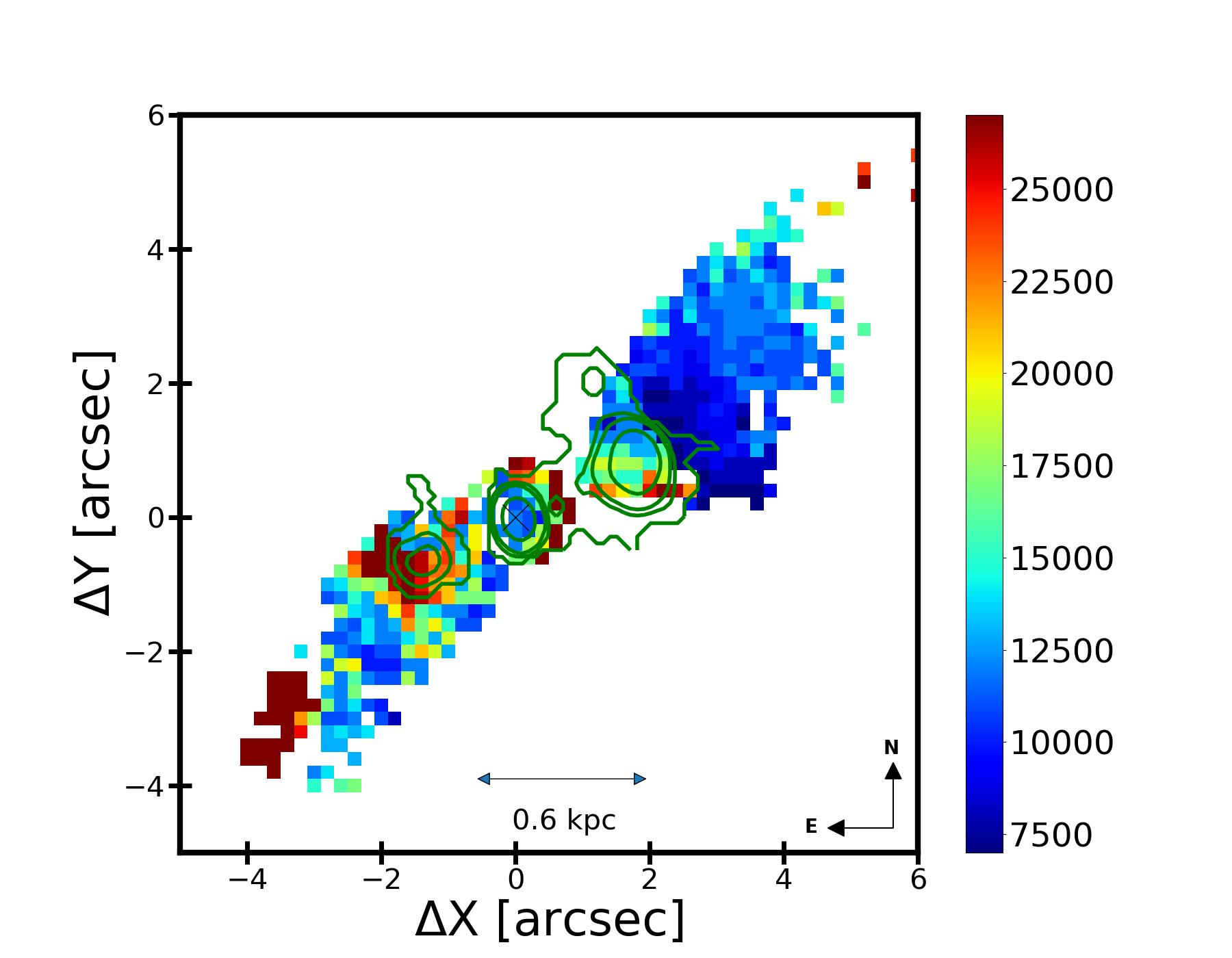}%[width=8.2\textwidth]
     \caption[Temperature map for IC\,5063.]{Temperature map for IC\,5063 (in Kelvin units) determined from the [\ion{S}{iii}]\,$\lambda\lambda$9068, +9531/$\lambda$6313 emission line flux ratio. The contours in green mark the radio emission region.}
     \label{fig:cap5temp}
\end{figure}

Figure \ref{fig:cap5temp} shows the map of the electron temperature, $T_{\rm e}$ in IC\,5063.  It could be determined mainly in the nuclear region and in and around the SE and NW radio lobes. In the southeast, the temperature is of the order of 20000\,K in the region emitting [\ion{Fe}{vii}], close to the radio lobe and $<$17000\,K outside it.  To the northwest, a mean temperature of 15000\,$\pm$\,3000\,K is derived inside the radio lobe region,  with minimum values reaching 7500~K. The largest values to the NW are observed at the edges of the [\ion{Fe}{vii}] emission, coinciding with the NW radio-lobe.  
The above values are, within uncertainties, in agreement with the electron temperature derived by \citet{HOLDEN_2022} using the [\ion{O}{iii}]~$\lambda$5007/$\lambda$4363 lines at the same locations. They find electron temperatures in the range
11500~K $< T_{\rm e} <$ 14000~K.  Very recently, \citet{dasyra_22} determined the temperature in IC\,5063 using the same MUSE data set as here but employing the [\ion{N}{ii}]~$\lambda\lambda$6548 +6583/[\ion{N}{ii}]~$\lambda$5755 emission line flux ratio. At the same locations shown in Figure~\ref{fig:cap5temp}, they derived electron temperatures that have minimum values similar to those obtained here ($\sim$ 7000 K). However their highest values tend to be lower than those derived by us, with maximum values reaching 12000~K (similar temperature was identified by \citet{HOLDEN_2022}).  We attribute this discrepancy to the fact that the [\ion{N}{ii}] lines are mostly produced by radiation-dominated clouds (see Sect.~\ref{sec:models}), associated to lower temperature gas. They would not be suitable to measure the temperature in high-density outflowing gas. It is important to mention that the gas density has little effect on the temperature. Had we used densities in the range 10$^{2-3}$~cm$^{-3}$ derived from the [\ion{S}{ii}] ratio by \citet{mingozzi_2019} would translate into differences of just $\sim$150~K at the most when T$_{\rm e} \sim 2 \times 10^4$ K. For smaller temperatures (i.e., 10$^4$~K), changing the gas density by one dex translates into differences of a few tens of K.

The results above indicate a close relationship between the increase in gas temperature  (See Fig.~\ref{fig:cap5temp}), the region of radio emission, and the emission of [\ion{Fe}{vii}] (see Fig.~\ref{fig:channels_ic5063}). It is also interesting that the highest temperature values are co-spatial to the regions of higher gas turbulence (Fig. ~\ref{fig:channels_ic5063}), higher FWHM values (Figure \ref{fig:cinematica_ic5063}), most notably in the NW hotspot than in the SE one, and higher gas density \citep{HOLDEN_2022}. Additionally, the fact that the coronal gas does not follows the galaxy rotation, showing velocities in excess of several hundred km\,s$^{-1}$, demonstrates that this gas is being dragged and linked directly with the \outs\ already detected at other wavelengths.

\section{Modelling the spectral line ratios}
\label{sec:models}

In the previous sections, we found that the gas ionisation in IC\,5063 increases as we approach the jet's axis, being the highest at the position of the AGN as well as very close to the radio-lobes, hundreds of parsecs away from the central engine. Indeed, at the NW and SE radio lobes, the [\ion{Fe}{vii}] emission is enhanced, reaching one tenth of its value at the nucleus.  A similar result was already reported in Circinus by \citet{fonseca-faria_2021}. They found peaks of [Fe\,{\sc vii}] emission, of the same order of intensity as that measured in the nucleus, at regions located hundreds of parsecs away from the AGN.  

According to \citet{ferguson1997}, in a photoionisation by the central source scenario, gas ionisation is expected to decrease when the distance to the AGN increases. From their models, coronal lines could not be formed at distances greater than 100\,pc from the central engine if the bolometric luminosity, $L_{\rm bol}$, is less than $10^{43.5}$\,erg\,s$^{-1}$.  \citet{ardila_2017} carried out CLOUDY simulations for NGC\,4388 ($L_{\rm bol}$=$1.26 \times 10^{44}$~\,erg\,s$^{-1}$). They found that the CLR was restricted to a few hundred parsecs from the AGN at most. This result can be easily extended to the case of IC\,5063, which has a very similar bolometric luminosity  ($L_{\rm bol}$=$10^{44}$~\,erg\,s$^{-1}$). It can be argued that gas with very low density ($n_{\rm e} < 50$~cm$^{-3}$) and photoionised by the central source may produce coronal emission thousand of parsecs away from the AGN. However, as we have already mentioned in Section~\ref{sec:gasden}, such values of electron density, particularly in the radio lobes or close to them, are not observed \citep{mingozzi_2019, HOLDEN_2022}. Thus,  photoionisation by the AGN alone is not able to explain the kiloparsec size of the observed CLR in this object. An additional source of gas ionisation  is likely producing a local increase in gas excitation along the direction where the [Fe\,{\sc vii}] is observed.   

In the light of the results gathered here, in this section we model the observed emission line spectra in IC\,5063 using the combined effect of photoinisation by the AGN and shocks, the latter likely produced by the passage of a radio jet accross the ISM.

\subsection{Calculation details}

We have modelled the spectra extracted from 10 selected regions, located at different distances from the AGN in the IFU cube. They were chosen as representative of the different environments found in the inner few kiloparsecs around the AGN. The size of each box was set to keep a signal/noise $>$ 3 in all the lines involved. For this reason,  its size varies accross the datacube, according to its location. Emission line fluxes were summed up and normalised relative to H$\beta$ within each individual box.

\begin{figure*}
\centering
\includegraphics[width=8.8cm]{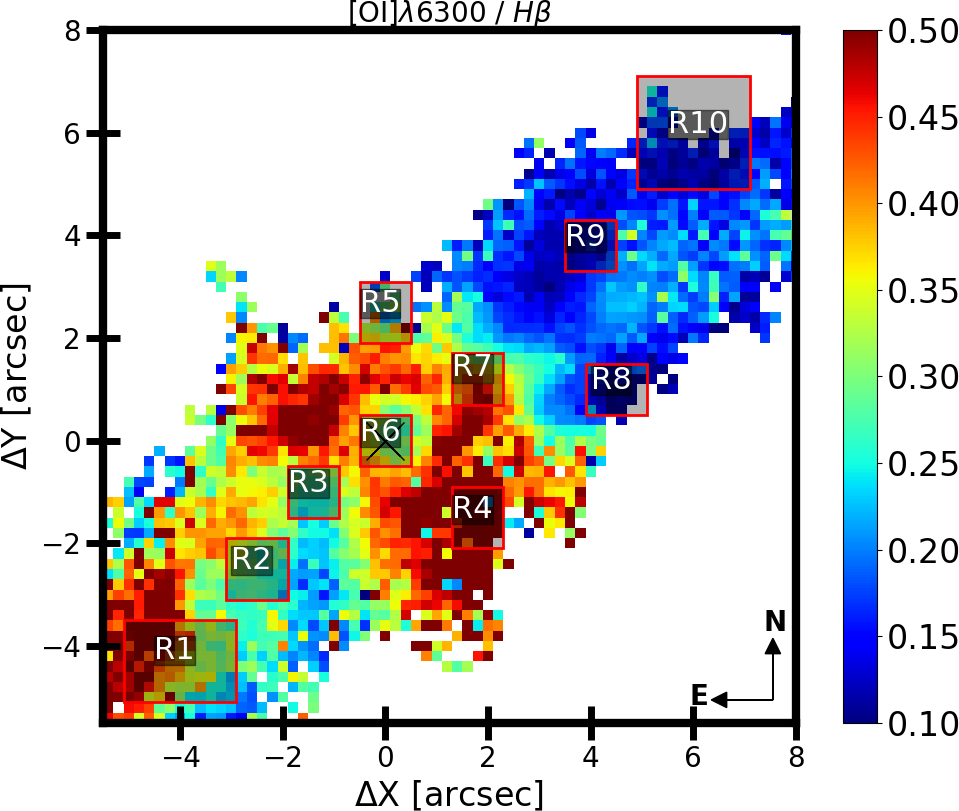} 
\includegraphics[width=8.8cm]{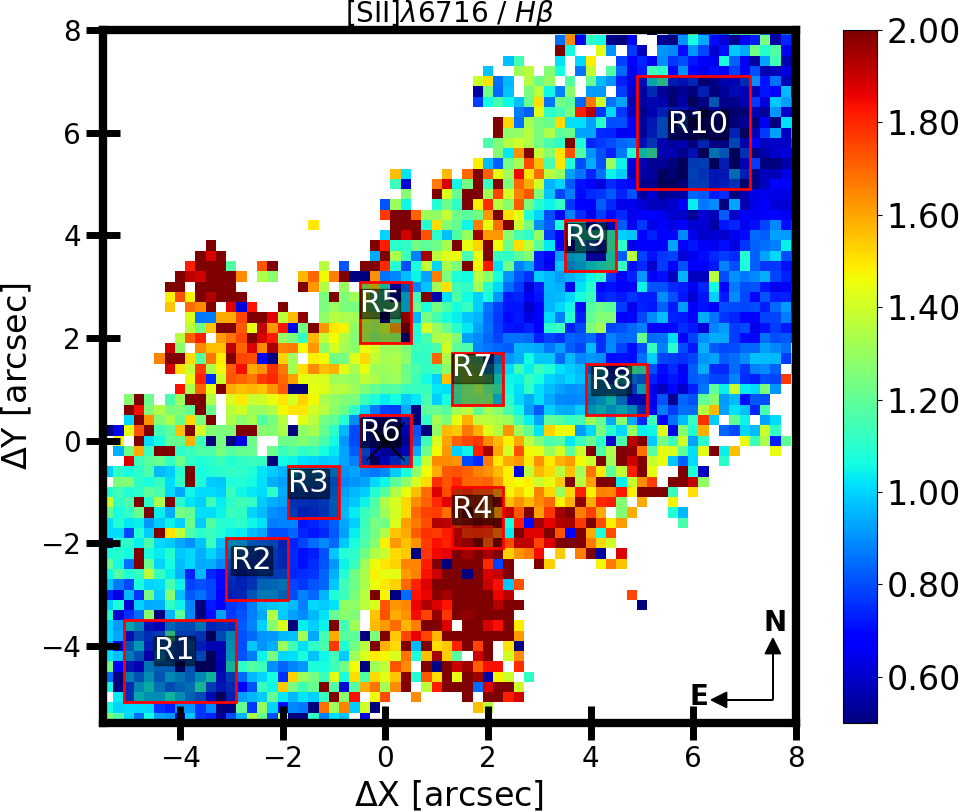}
\includegraphics[width=9cm]{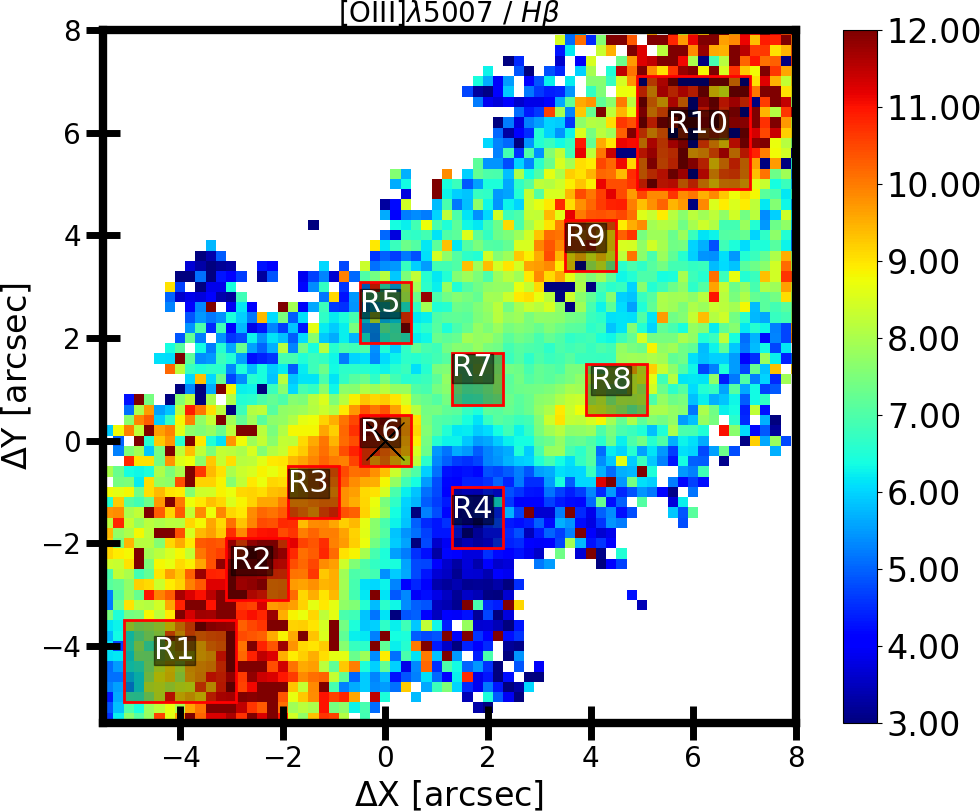}
\includegraphics[width=8.7cm]{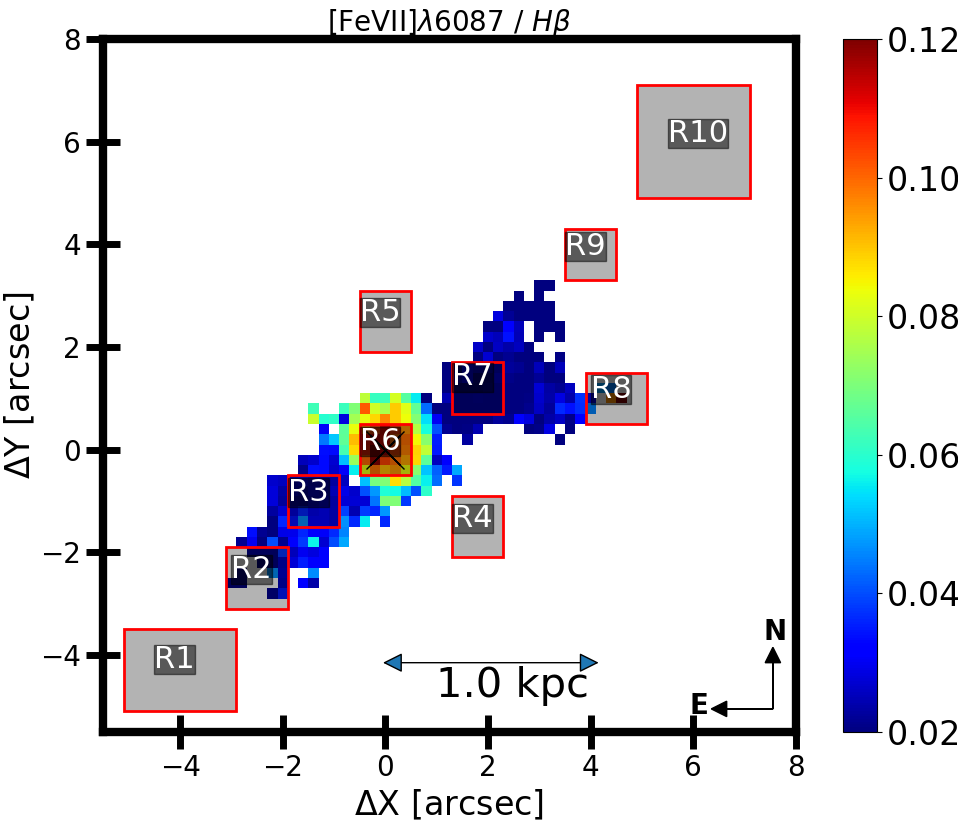}

\caption{
Emission line flux ratios 
[O\,{\sc i}]~$\lambda$6300/$H\beta$ (top left),
[S\,{\sc ii}]~$\lambda$6716/$H\beta$ (top right),
[O\,{\sc iii}]~$\lambda$5007/$H\beta$ (bottom left),
and [Fe\,{\sc vii}]~$\lambda$6087/$H\beta$ (bottom right).
Ten regions, marked in red, where selected as representative of the ionisation gas. Region R6 coincides with the position of the AGN while regions R3 and R7 are set at the position of the SE and NW radio lobes, respectively. Regions R1 to R3 and R9-R10 are located along the axis of the ionisation cone while regions R4, R5, and R8 are positioned in the inter-cone region or very close to it. The two bottom panels are zoom-in versions of the maps presented in Fig.~\ref{fig:razaolinhasgalaxias}. The scale bar, shown in the bottom right panel, is the same for all maps in the Figure.}
\label{fig:ratios_to_hbeta}
\end{figure*}

Figure \ref{fig:ratios_to_hbeta} shows maps of the emission line flux ratios
[O\,{\sc i}]~$\lambda$6300/H$\beta$ (top left),
[S\,{\sc ii}]~$\lambda$6716/H$\beta$ (top right),
[O\,{\sc iii}]~$\lambda$5007/H$\beta$ (bottom left),
and [Fe\,{\sc vii}]~$\lambda$6087/H$\beta$ (bottom right), with the different regions overlaid.   Region R6 is centred at the AGN; regions R3 and R7 are at the position of the SE and NW radio lobes, respectively.
Regions R1 to R3 and R9 to R10 are located along the ionisation cone while regions R4, R5, and R8 are positioned in the inter-cone region or very close to it.

In order to reproduce the average line ratios in the different regions we used the photoionisation code {\sc suma}~\citep{Contini_2001}. It combines the effect of shocks (i.e. produced by the passage of a jet) and photoionization by the central source.
As explained in \citet{fonseca-faria_2021}, the main input parameters are those which lead  to the calculations of  line and continuum fluxes that best matches the observed ones.
They account for photoionisation and heating by  primary and secondary  radiation and for collision process due to shocks.
The input parameters such as  the shock velocity \Vs, the atomic preshock density \n0 and the preshock
magnetic field \B0 (for all models  \B0~=~10$^{-4}$ Gauss is adopted) define the hydrodynamical field.
They  are combined in the compression equation \citep{cox_1972} which is resolved within  each slab of the gas in order to obtain the density profile  throughout the emitting clouds.
For an AGN, the  photoionising radiation is the power-law radiation   flux  from the active centre $F$,  in number of photons cm$^{-2}$ s$^{-1}$ eV$^{-1}$ at the Lyman limit
   and  spectral indices  $\alpha_{UV}$=-1.5 and $\alpha_X$=-0.7.

The primary radiation source
does not depend on the host physical condition but it affects the surrounding gas.   Each  region  is not considered
as a unique cloud, but as a  sequence of plane-parallel slabs (up to 300) with different geometrical thickness
calculated automatically following the temperature gradient.
The secondary diffuse radiation is emitted from the slabs of
gas heated  by the radiation flux reaching the gas and by the shock.
Primary and secondary radiation are calculated by radiation transfer.

The calculations initiate at the shock front where the gas is compressed and  adiabatically thermalised,
reaching a maximum temperature in the immediate post-shock region  $T$\, $\sim$~1.5$\times 10^5$ (\Vs/100 \kms)$^2$ (K).
$T$ decreases downstream  leading to recombination.  The cooling rate is calculated in each slab.
The line and continuum emitting  regions throughout the galaxy cover  an ensemble of fragmented clouds.
The geometrical thickness $D$ of the clouds is  an input parameter  which is  calculated
consistently with the physical conditions and element abundances of the emitting gas.
The fractional abundances of the ions are calculated resolving the ionisation equations
for each element (H, He, C, N, O, Ne, Mg, Si, S, Ar, Cl, Fe) in each ionisation level.
Then, the calculated line ratios, integrated throughout the cloud geometrical width, are compared with the
observed ones. The calculation process is repeated
changing  the input parameters until the observed data are reproduced by the model results,  at maximum
within 10-20 percent
for the strongest line ratios and within 50~percent for the weakest ones.

\subsection{Physical conditions and relative element abundances in the different regions}

The set of input parameter values that produces calculated  to  observed line ratios closest to unity is chosen as the best model.

In Tables \ref{tab:m1} and \ref{tab:m2}, the observed line ratios (R1-R10) are compared with model results in rows 1-12.  In row 13 the calculated \Hb\ fluxes appear for each model 
 followed by the observed \Hb\ fluxes (\Hb$_{o}$) in row  14. The physical input parameters of the models are given in rows 15-18 followed by the relative abundances of the elements
adopted by the models in rows 19-22. The other element abundances  are assumed solar \citep{asplund_2009}.

The physical conditions in the nuclear and extended regions in IC\,5063 are such as to allow to reproduce
the whole spectrum from  different  areas  by  single models. 
In particular, the geometrical thickness of the clouds $D$ are large enough to contain well defined shock dominated 
(SD) and radiation dominated (RD) regions.
The models   reproduce the high ionisation-level line ratios
(e.g. [\ion{Fe}{vii}]/\Hb\ and [\ion{Fe}{x}]/\Hb) as well as the low ionisation ones such as [\ion{N}{ii}]/\Hb\ and the intermediate ones
(e.g.  [\ion{O}{iii}]/\Hb) within the accepted error. Notice that at some regions, SUMA predicts very small [\ion{Fe}{vii}]/\Hb\ ratios that, in practice, are very difficult to measure due to S/N issues. For instance, in regions R1 and R5, [\ion{Fe}{vii}] is not detected while the models predict values of 0.009 in that line (normalised to H$\beta$). In such cases we may safely assume that there is a good agreement between the predicted and observed values because we are not able to detect lines that are fainter than 1\% of the H$\beta$ intensity.

\begin{table*}
\centering
\caption{Calculated emission line flux ratios ({\sc{SUMA}} outputs) and  observed ratios for regions R1 to R5.  
}\label{tab:m1}
\begin{tabular}{lcccccccccccccc} \hline  \hline
\                     &R1 &R1   &R2  &R2                 &R3  &R3  &R4 &R4  &R5 &R5  \\ 
    &observed  &calculated &observed  &calculated        &observed  &calculated  &observed  &calculated &observed  &calculated                   \\ \hline
\ [\ion{O}{iii}]5007~/~\Hb\  &9.13 &8.8   &11.23 &10.0    &9.94 &9.4  &4.19&4.0  &7.17&7.30 \\
\ [\ion{Fe}{vii}]6087~/~\Hb\  &-   &0.009 &0.02 &0.011    &0.03&0.018 &-   &0.013&-   &0.009 \\
\  [\ion{O}{i}]6300~/~\Hb\  &0.40  &0.29  &0.28 &0.24     &0.28&0.26 &0.54 &0.4 &0.41&0.59 \\
\ [\ion{S}{iii}]6312~/~\Hb\  &0.21 &0.043 &0.03 &0.04     &0.04&0.04&-   &0.02 &-   &0.038\\
\ [\ion{Fe}{x}]6375~/~\Hb\  &-     &-     &-    &-        &0.04 &0.033  &-   &-    &-   &-    \\
\  \Ha\,6563~/~\Hb\         &3.19  &2.9   &3.11 &2.9      &3.09&3.09  &3.22&2.95 &3.27&2.93 \\
\ [\ion{N}{ii}]6583~/~\Hb\  &2.22  &2.0   &1.87 &2.19     &2.32&2.0  &3.42&3.1  &2.89&2.7  \\
\ [\ion{S}{ii}]6716~/~\Hb\  &0.81  &0.82  &0.84 &0.73     &0.85&0.72 &1.86&0.6  &1.23&1.0  \\
\ [\ion{O}{ii}]7320~/~\Hb\  &0.05  &0.1   &0.04 &0.1      &0.06&0.08 &-   &0.08 &-   &0.09 \\
\ [\ion{S}{xii}]7611~/~\Hb\  &- &- & - &-          &-    &-       &-    &- &-                &- \\
\ [\ion{Fe}{xi}]7892~/~\Hb\  &- &- & - &-          &-    &-       &-    &- &-                &- \\
\ [\ion{S}{iii}]9069~/~\Hb\  &0.34 &0.9   &0.41 &0.8      &0.45&0.6  &-   &0.5  &-   &0.8  \\
\ \Hb$_c$  $^1$                  &-    &0.06  &-  &0.043      &-   &0.043&-   &0.05 &-   &0.06 \\
 \ \Hb$_o$ $^2$               &2.94  &-   &7.44 &-        &42.4  &-    &1.46&-    &2.14&-    \\
\ \Vs $^3$                    &-   &420   &-    &380      &-   &380  &-   &500  &-   &390  \\
\ \n0 $^4$                    &-   &160   &-    &140      &-   &140  &-   &150  &-   &170  \\
\ $D$ $^5$                    &-   &2.48  &-    &2.9      &-   &2.9  &-   &2.5  &-   &3.5  \\
\ $F$ $^6$                    &-   &3.2   &-    &2.6       &-   &2.6  &-   &1.8  &-   &3.   \\
\ (N/H) $^7$                  &-   &0.8   &-    &0.8       &-   &0.9  &-   &1.3  &-   &1.   \\
\ (O/H) $^7$                  &-   &6.6   &-    &6.4       &-   &6.4  &-   &6.7  &-   &6.0  \\
\ (S/H) $^7$                  &-   &0.2   &-    &0.16      &-   &0.16 &-   &0.16 &-   &0.19 \\
\ (Fe/H) $^7$                 &-   &0.32  &-   &0.32      &-   &0.42 &-   &0.32 &-   &0.32 \\ 
\ r$_c$ $^8$                 &-    &0.77  &-   &1.45      &-   & 3.45 &-   &0.59 &-   &0.66\\ 
\ r$_{c}\times$~\ff~$^9$        &-   &0.77  &-    &0.87      &-   &0.45 &-   &0.59 &-   &0.66  \\ 
\  r$_o$~$^{10}$                &-   &1.40  &-    &0.92      &-   &0.44 &-   &0.68 &-   &0.68  \\
\  \ff~$^{11}$                &-   & 1.0  &-    &0.60      &-   &0.13 &-   &1.0  &-   &1.0   \\ \hline

\end{tabular}
         
             $^{1}$ Calculated flux at the galaxy in \erg; $^2$ observed flux at Earth in 10$^{-15}$ \erg; $^3$ calculated
shock velocity in \kms;  $^4$ calculated preshock density in \cm3; $^5$ calculated geometrical depth of the gaseous slab in 10$^{16}$ cm; $^6$ calculated primary flux from the AGN in 10$^{10}$ photons cm$^{-2}$ s$^{-1}$ eV$^{-1}$ 
at the Lyman limit; $^7$ element relative abundance in 10$^{-4}$ units;  
$^8$ calculated distance from the AGN (coinciding with R6) in kpc; $^9$  r$_c$ multiplied by the filling factor; $^{10}$ observed distance in kpc measured in Fig. \ref{fig:ratios_to_hbeta}; $^{11}$ filling factor
        
\end{table*}

\begin{table*}
\centering
\caption{Same as Table~\ref{tab:m1} for regions R6 to R10.} \label{tab:m2}
\begin{tabular}{lcccccccccccccc} \hline  \hline
\             &R6  &R6  &R7  &R7  &R8 &R8  &R9 &R9  &R10 &R10  \\
    &observed  &calculated &observed  &calculated &observed  &calculated  &observed  &calculated &observed  &calculated                   \\ \hline
\ [\ion{O}{iii}]5007~/~\Hb\   &10.27& 10.26  &7.40  &7.24    &8.10 &8.4    &9.14&8.8  &9.88&10.0\\
\ [\ion{Fe}{vii}]6087~/~\Hb\  &0.1  &0.247   &0.02 &0.018    &0.07 &0.06   &-   &0.01 &-   &0.01\\
\ [\ion{O}{i}]6300~/~\Hb\     &0.3  &0.15        &0.40  &0.4     &0.15 &0.02   &0.14&0.13 &0.14&0.23\\
\ [\ion{S}{iii}]6312~/~\Hb\  & 0.10 &0.046    &0.04  &0.03     &0.05 &0.05    &0.03&0.046&-   &0.04\\
\ [\ion{Fe}{x}]6375~/~\Hb\  &0.07 &0.067    &0.02  &0.02   &-   &0.016   &-   &-    &-   &0.002\\
\  \Ha6563~/~\Hb\           & 3.12&3.22      &3.11  &2.91     &3.16 &2.9     &3.17&2.88 &3.14&2.9\\
\ [\ion{N}{ii}]6583~/~\Hb\  &1.96 &1.98      &2.37  &2.4     &1.9  &1.8     &1.59&1.56 &1.31&1.50\\
\ [\ion{S}{ii}]6716~/~\Hb\  &0.65 &0.30      &1.19  &0.5     &0.92 &0.7     &0.95&0.5  &0.67&0.71\\
\  [\ion{O}{ii}]7320~/~\Hb\  &0.07 &0.49     &0.08  &0.11    &-    &0.13    &0.04&0.13 &-   &0.11 \\
\  [\ion{S}{xii}]7611~/~\Hb\  &0.010 &0.004  & - &-          &-    &-       &-    &- &-                &- \\
\  [\ion{Fe}{xi}]7892~/~\Hb\  &0.030 &0.012  & - &-          &-    &-       &-    &- &-                &- \\
\ [\ion{S}{iii}]9069~/~\Hb\  &0.38 &0.62      &0.40   &0.70   &0.46 &1.1     &0.32&0.7  &0.23&0.8  \\
\ \Hb$_c$ $^1$   &-   &0.002      &-            &0.01   &-     &0.011    &-    & 0.08&-   &0.055\\ 
\ \Hb$_o$ $^2$   &103   &-    &43.2         &-      &1.19 &-    &6.48 &-   &4.79 & - \\
\ \Vs $^3$   &-    &450      &-    &600 &-   &450   &-   & 540 &-   &400  \\
\ \n0 $^4$    &-    &170      &-    &150 &-   &170   &-   & 150 &-   &150  \\
\ $D$ $^5$    &-    &0.85     &-    &3.6 &-   &0.95  &-   & 2.34&-   &2.7  \\
\ $F$ $^6$    &-    &0.7      &-    &5.3 &-   &1     &-   & 3.8 &-   &3.   \\
\ (N/H) $^7$  &-    &0.7      &-    &1.0 &-   &0.7   &-   & 0.6 &-   &0.6  \\
\ (O/H) $^7$  &-    &6.3      &-    &6.2 &-   &6.6   &-   & 6.3 &-   &6.2  \\
\ (S/H) $^7$  &-    &0.10     &-    &0.10&-   &0.16  &-   & 0.16&-   &0.16 \\
\ (Fe/H) $^7$ &-    &0.15     &-    &0.42&-   &0.32  &-   &0.32 &-   &0.32 \\ 
\   r$_c$  $^8$&-    &25    &-    &2.7 &-   & 1.14 &-   &0.99 &-   &1.02\\ 
\ r$_{c}\times$~\ff~$^9$ &- &0.1  &-&0.57 &-   & 1.14 &-   &0.99 &-   &1.02\\ 
\  r$_o$~$^{10}$ &-     &0.1    &-    &0.6 &-   &1.2   &-   &1.41 &-   &2.20\\  
\  \ff ~ $^{11}$ &-   &0.004&-    &0.21&-     &1.0   &-   & 1.0 &-   &1.0 \\  \hline
\end{tabular}

            $^1$ Calculated flux at the galaxy in \erg; $^2$ observed flux at Earth in 10$^{-15}$ \erg;  $^3$ calculated
shock velocity in \kms;  $^4$ calculated preshock density in \cm3; $^5$ calculated geometrical depth of the 
gaseous slab in 10$^{16}$ cm; $^6$ calculated primary flux from the AGN in 10$^{10}$ photons cm$^{-2}$ s$^{-1}$ eV$^{-1}$ 
at the Lyman limit; $^7$ element relative abundance in 10$^{-4}$ units;  
$^8$ calculated distance from the AGN (coinciding with R6) in kpc; 
$^9$  r$_c$ multiplied by the filling factor; 
$^{10}$ observed distance in kpc measured in Fig. \ref{fig:ratios_to_hbeta}; 
$^{11}$ filling factor
        
\end{table*}

\begin{figure*}
\centering 
\includegraphics[width=8.0cm]{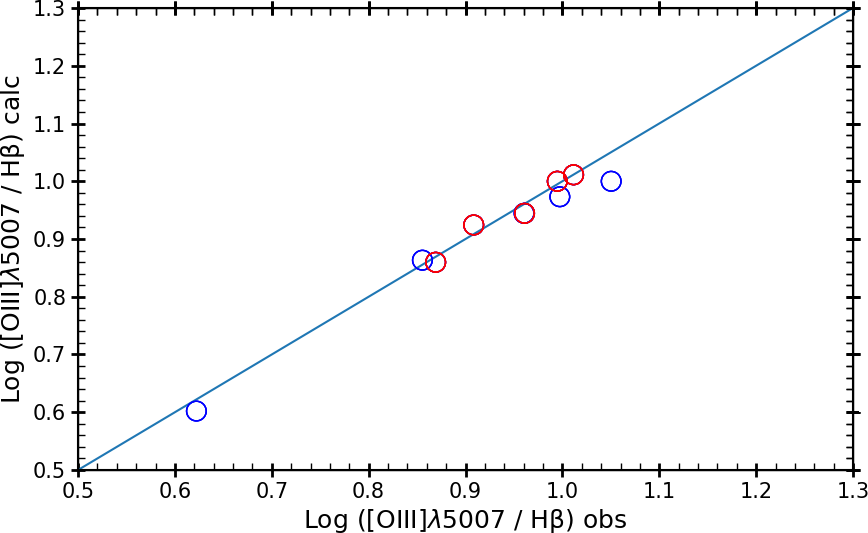}
\includegraphics[width=8.0cm]{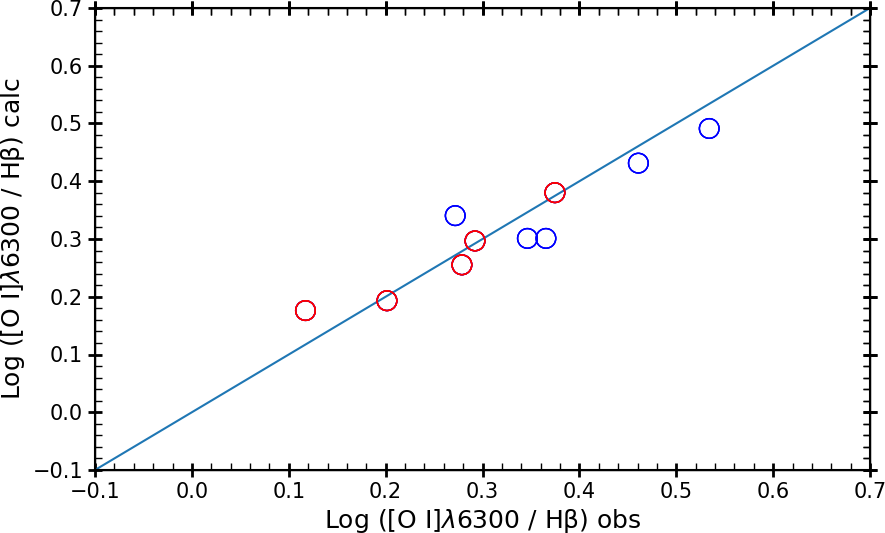}
\includegraphics[width=8.0cm]{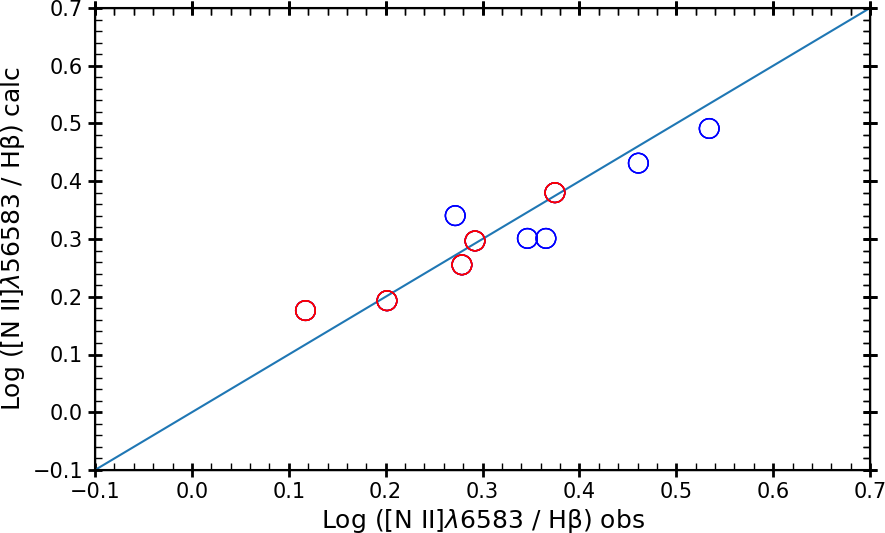}
\includegraphics[width=8.0cm]{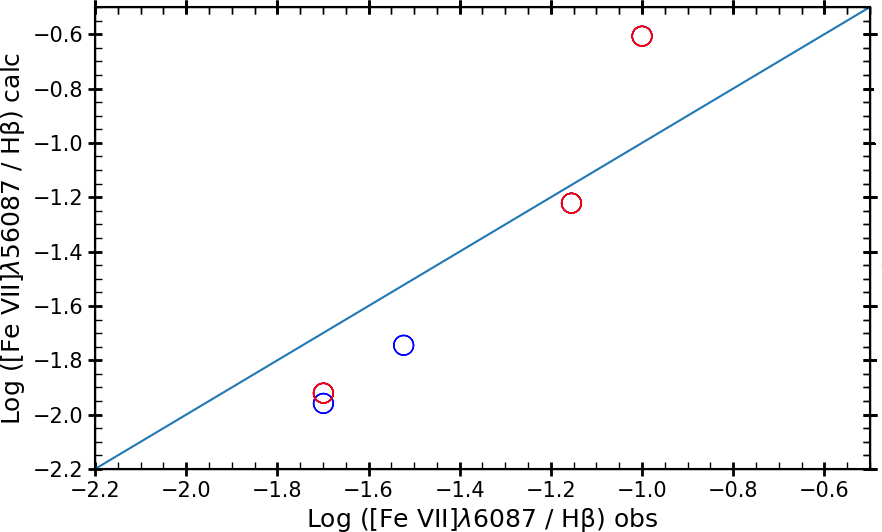}

\caption{ Comparison of calculated with observed line ratios.
Blue circles: regions R1, R2, R3, R4 and R5.
Red  circles: regions R6, R7, R8, R9 and R10.}
%\label{fig1}
\label{fig_model_1}
\end{figure*}

Tables \ref{tab:m1}-\ref{tab:m2}  and Fig.\ref{fig_model_1} show that the most significant  calculated line ratios satisfactorily reproduce the observations,
in particular [\ion{O}{iii}]/\Hb\ and [\ion{N}{ii}]/\Hb, which are the strongest ones. We notice that the ratios involving the coronal lines [\ion{Fe}{vii}] and [\ion{Fe}{x}] are also nicely reproduced in the nucleus as well as in the extended region.
 
In regions R1, R6, and R8, the  [\ion{O}{i}]/\Hb\ line ratios are underpredicted by the models because the  observed 
[\ion{O}{i}]/\Hb\ is likely enhanced by the contribution of the diffuse ISM \citep{mannucci21, belfiore22}. Oxygen is recombined in the cool ISM and can emit a not negligible [\ion{O}{i}] line flux. The calculated [\ion{S}{ii}]/\Hb\ line ratios are lower than the observed ones in the same regions
where the calculated [\ion{O}{i}]/\Hb\ are lower than observed. The first sulphur ionization potential (10.36 eV)
is lower than that of H (13.6 eV)  and of O (13.6 eV). Therefore we can assume that
$\text{S}^{ \text{\scalebox{1.0}{\!\,+}} }$  behaves  like O$^0$, suggesting
that the observed [\ion{S}{ii}] lines may contain as well  a non negligible contribution from the diffuse ISM.

\subsection{Relative contributions of shocks and AGN radiation to the lines}

\begin{table*}
\centering
\caption{Percent contributions to the most significant lines  from the shock dominated (SD) and radiation dominated (RD) clouds predicted by the models in regions
R1-R5 }
\label{tab:3}
\begin{tabular}{lcccccccccccccc} \hline  \hline
\                         & SD1  &  RD1   &SD2   & RD2   & SD3   & RD3   & SD4 &RD4  & SD5   & RD5   \\ \hline
\  \Hb\                   &0.3   & 99.7   &0.21  &99.8   & 1.3   &98.7   & 13.0 &87.0  &3.2    &96.8&\\
\ [\ion{O}{iii}]5007      &1.2   & 92.8   &1.33  &98.7   & 1.5    &98.5   &3.3  &96.7 &1.37   &98.6  \\
\  [\ion{Fe}{vii}]6087    &98.1  & 1.9    &95.9  &4.08   & 93.0  &7.0    &100  &0.0  &98.    &1.6  \\
\  [\ion{Fe}{x}]6375      &97.5  & 2.5    &99.8  &0.2    & 99.0   &1.0    &97.6 &0.01 &100   &0.0 \\
\  [\ion{N}{ii}]6583      &0.7   & 99.3   &0.6   &99.4   & 1.7   &98.3   &3.9 &96.1 &1.5    &98.5\\
\ [ \ion{O}{ii}]7320      &4      &  96   &5.6   &94.4  &9.3    &90.7   &5.9    &94.1 &8.5  &91.5 \\
\ T$_0$$^1$               &255  & 1.03   &209  &1.1    & 210  &1.04   &360 &0.94 &220   &1.0\\ \hline
\end{tabular}

\ $^1$ T$_0$ (in 10$^4$K) is the gas temperature at the edges of the cloud (see Fig. \ref{fig_model_1}).

\end{table*}

\begin{table*}
\centering
\caption{Same as Table~\ref{tab:3} for regions R6-R10.}
\label{tab:4}
\begin{tabular}{lcccccccccccccc} \hline  \hline

\  line             & SD6 & RD6 &SD7 & RD7  &SD8  & RD8 & SD9 &RD9 & SD10& RD10\\ \hline
\ \Hb \              & 24.9 & 75.1 &0.2 & 99.8  &0.6  &99.4 &0.2  &99.8 &0.2  &99.8 \\
\ [\ion{O}{iii}]5007  &43.0 & 57.0  &0.4 & 99.6 &2.8  &97.2 &2.8  &97.  &1.7  &98.3 \\
\ [\ion{Fe}{vii}]6087 & 100 & 0.0 &100 & 0.0  &99.0 &1.0  &99.8 &0.2  &98.1 &1.9  \\
\ [\ion{Fe}{x}]6375   & 100 & 0.0   &100& 0.0  &100 &0.0  &100 &0.0  &100 &0.0  \\
\ [\ion{N}{ii}]6583   &32.6  &67.4  &0.1 &99.9  &3.5  &96.5 &0.2  &99.8 &0.5  &99.5\\
\ [\ion{O}{ii}]7320   &71.4  &28.6  &0.2 &99.8  &0.8  &99.2 &2.5  &97.5 &6.2  &93.8 \\
\ T$_0$$^1$           &293 &0.96  &520 &1.07  &293 &0.94 &420 &1.0  &230 &1.0  \\ \hline

\end{tabular}

\ $^1$ T$_0$ (in 10$^4$K) is the gas temperature at the edges of the cloud (see Fig. \ref{fig_model_1})

\end{table*}

\begin{figure*}
\centering

\includegraphics[width=8.8cm]{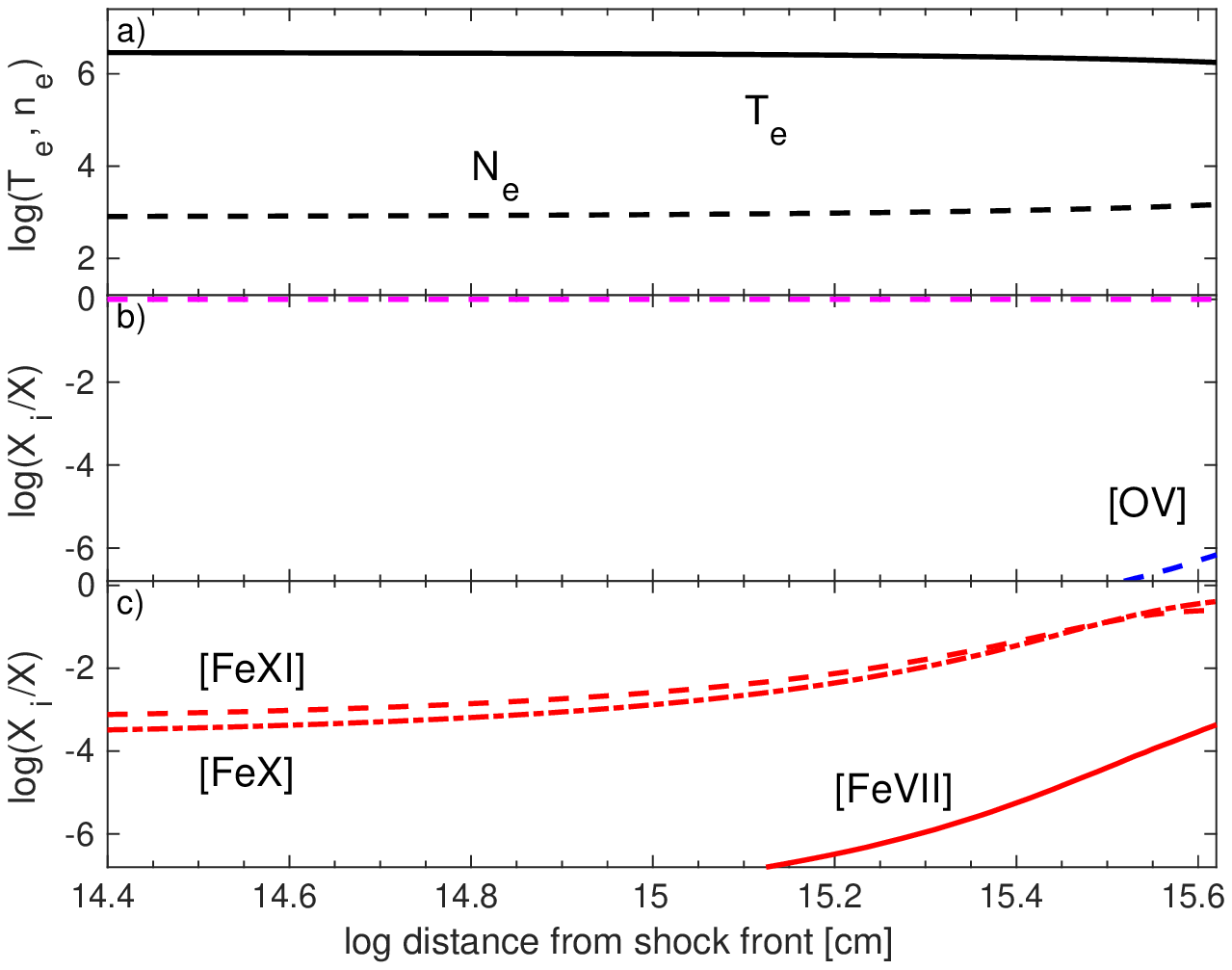}
\includegraphics[width=8.8cm]{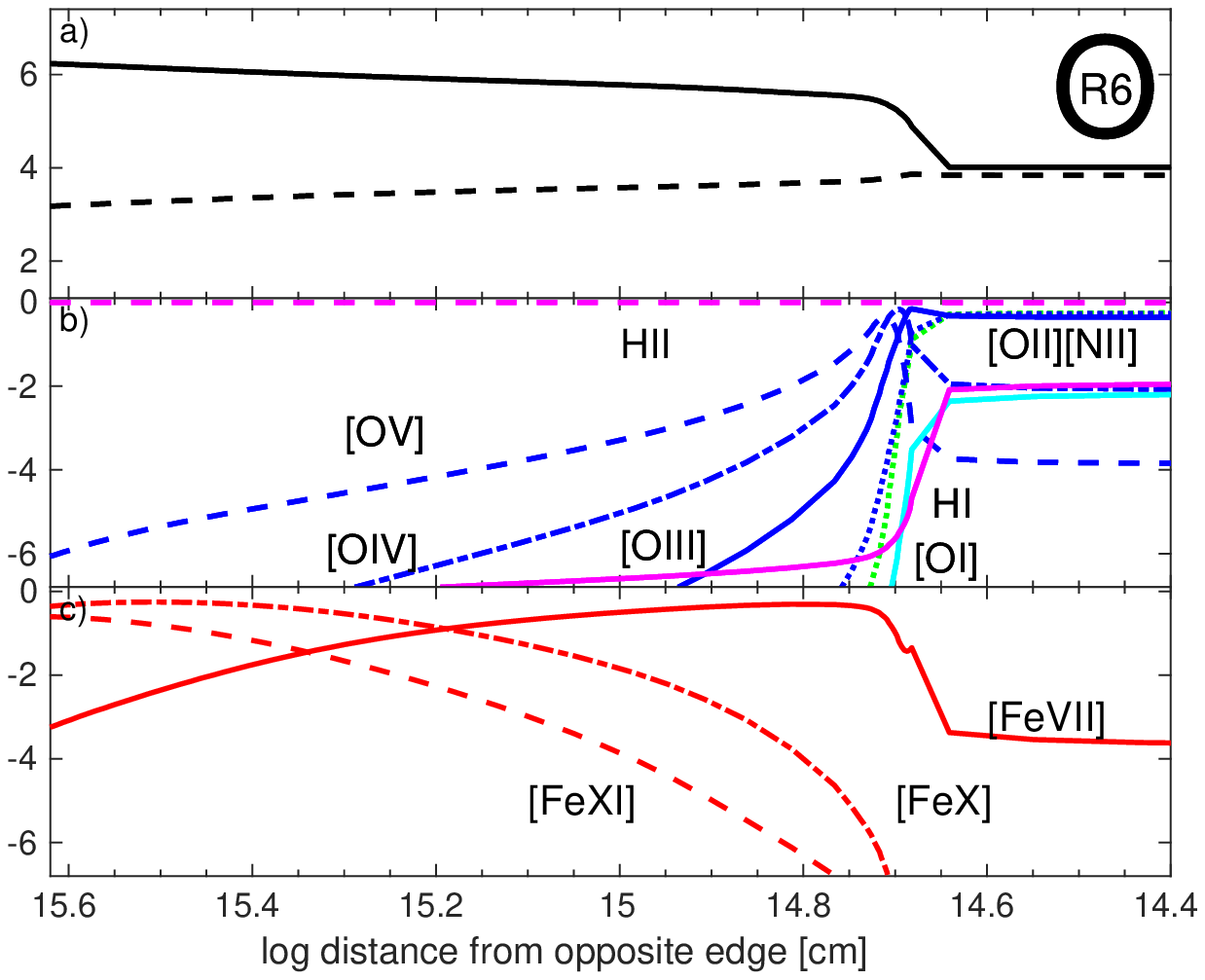}
\includegraphics[width=8.8cm]{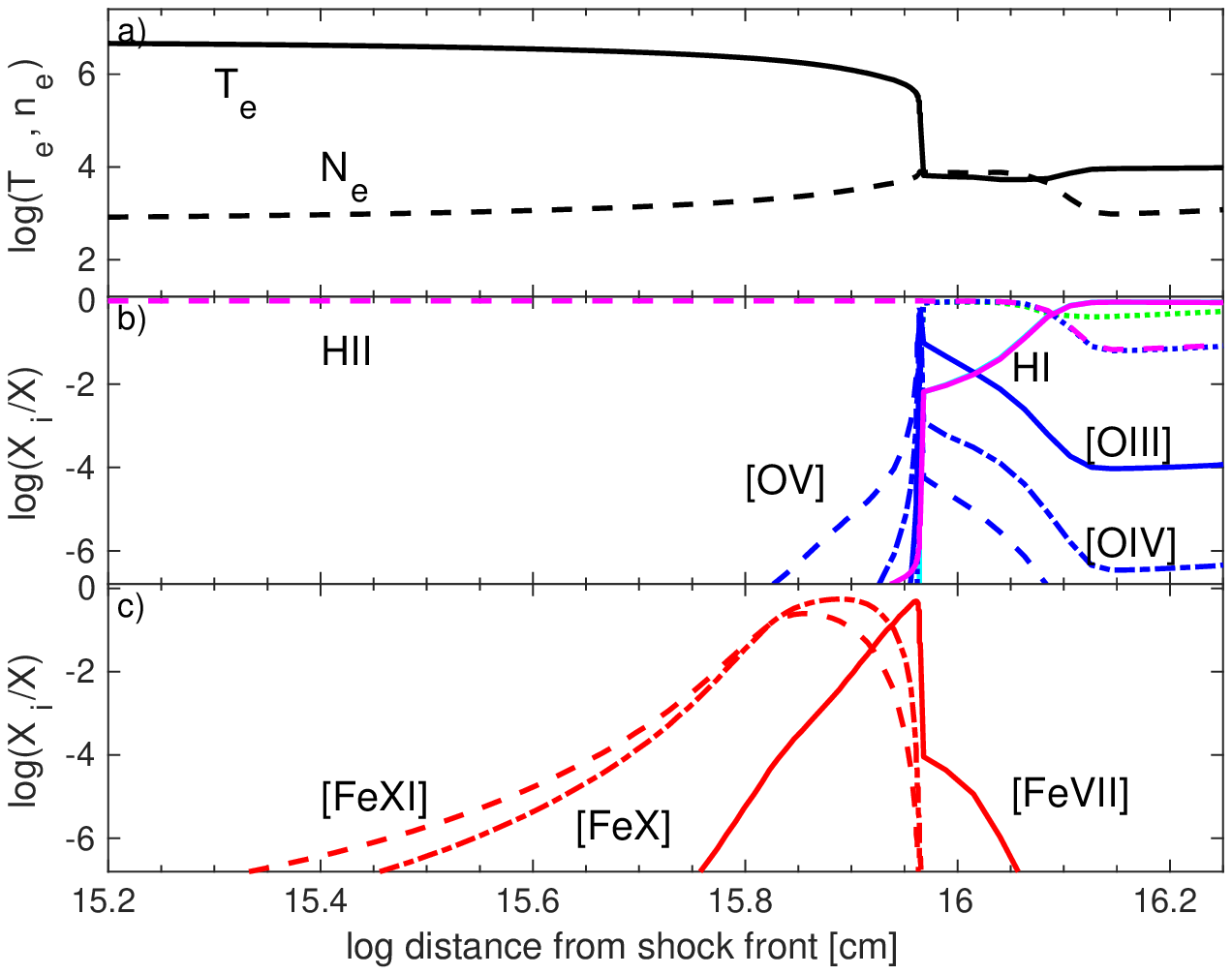}
\includegraphics[width=8.8cm]{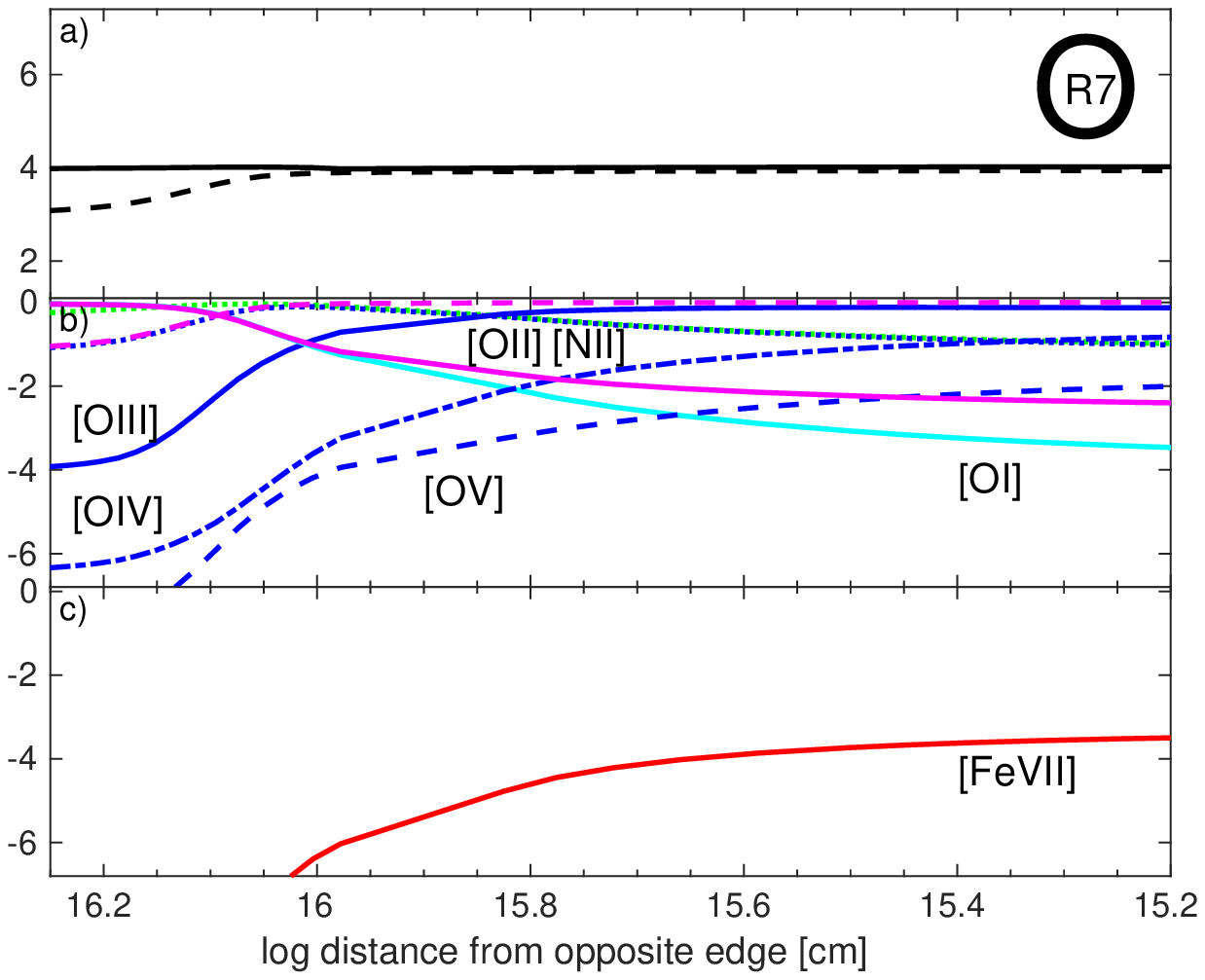}
\caption{\Te, \Ne ~and ion profiles through  regions R6 (top panels) and R7 (bottom panels) clouds.
The clouds are divided in two contiguous halves which are  displayed by the left and right panels.
In the left panel
the shock front is on the left and the X-axis scale is logarithmic. In the right panel
the right edge is reached by the flux from the active centre. The X-axis scale is logarithmic
in reverse in order to have the same detailed view of the cloud edges.
Black lines: \Te~(solid),  \Ne ~(dashed);
blue lines refer to ionized oxygen: [\ion{O}{v}] (dashed), [\ion{O}{iv}]  (dash-dotted), [\ion{O}{iii}]  (solid), [\ion{O}{ii}]  (dotted); cyan: [\ion{O}{i}]  (solid);  green: [\ion{N}{ii}]  (dotted); magenta: [\ion{H}{i}]  (solid), [\ion{H}{ii}]  (dashed);
red: [\ion{Fe}{xi}]  (dashed), [\ion{Fe}{x}]  (dash-dotted), [\ion{Fe}{vii}]  (solid).
	}
 \label{fig_abundancia1}
\end{figure*}

\begin{figure*}
\centering
\includegraphics[width=8.8cm]{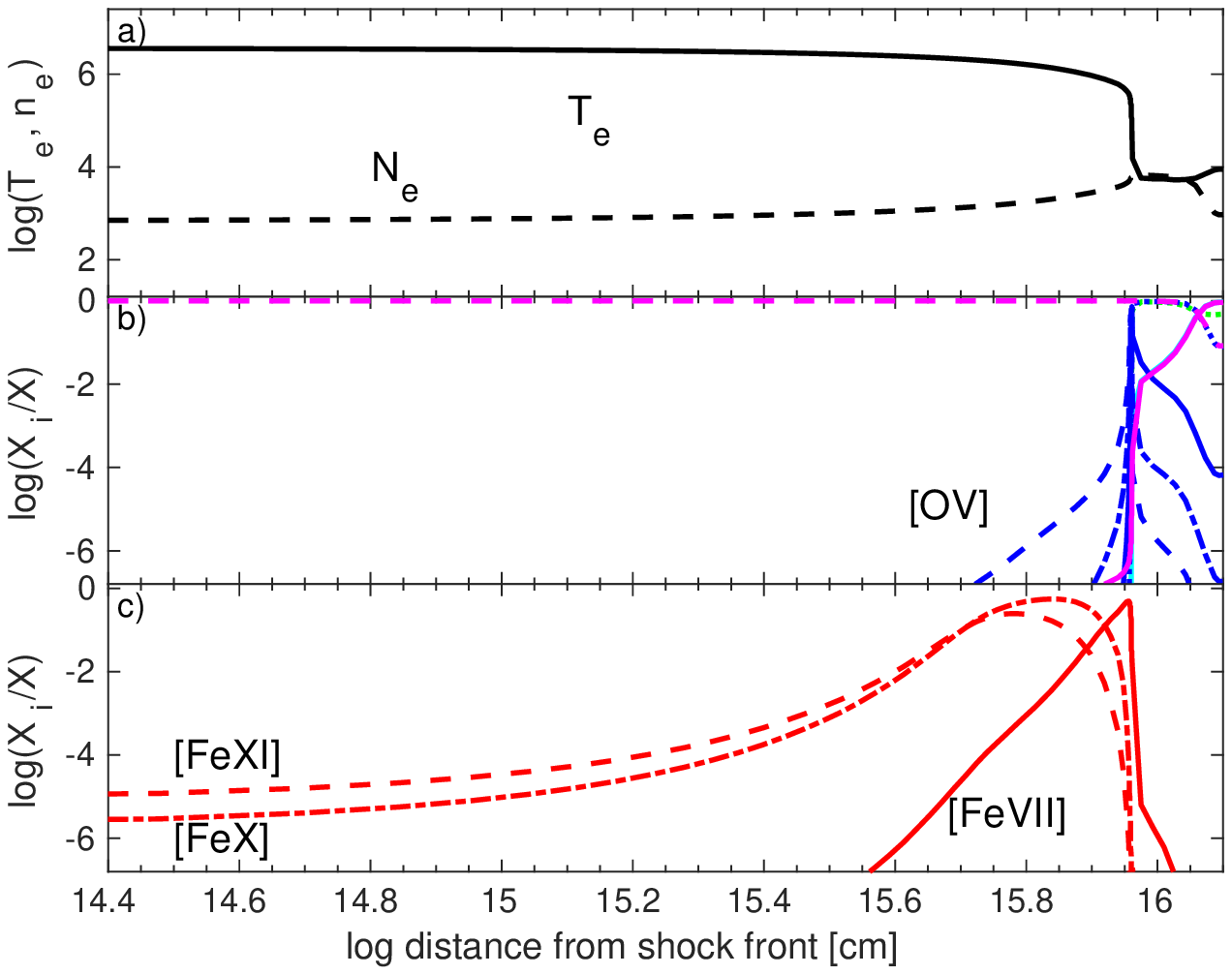}
\includegraphics[width=8.8cm]{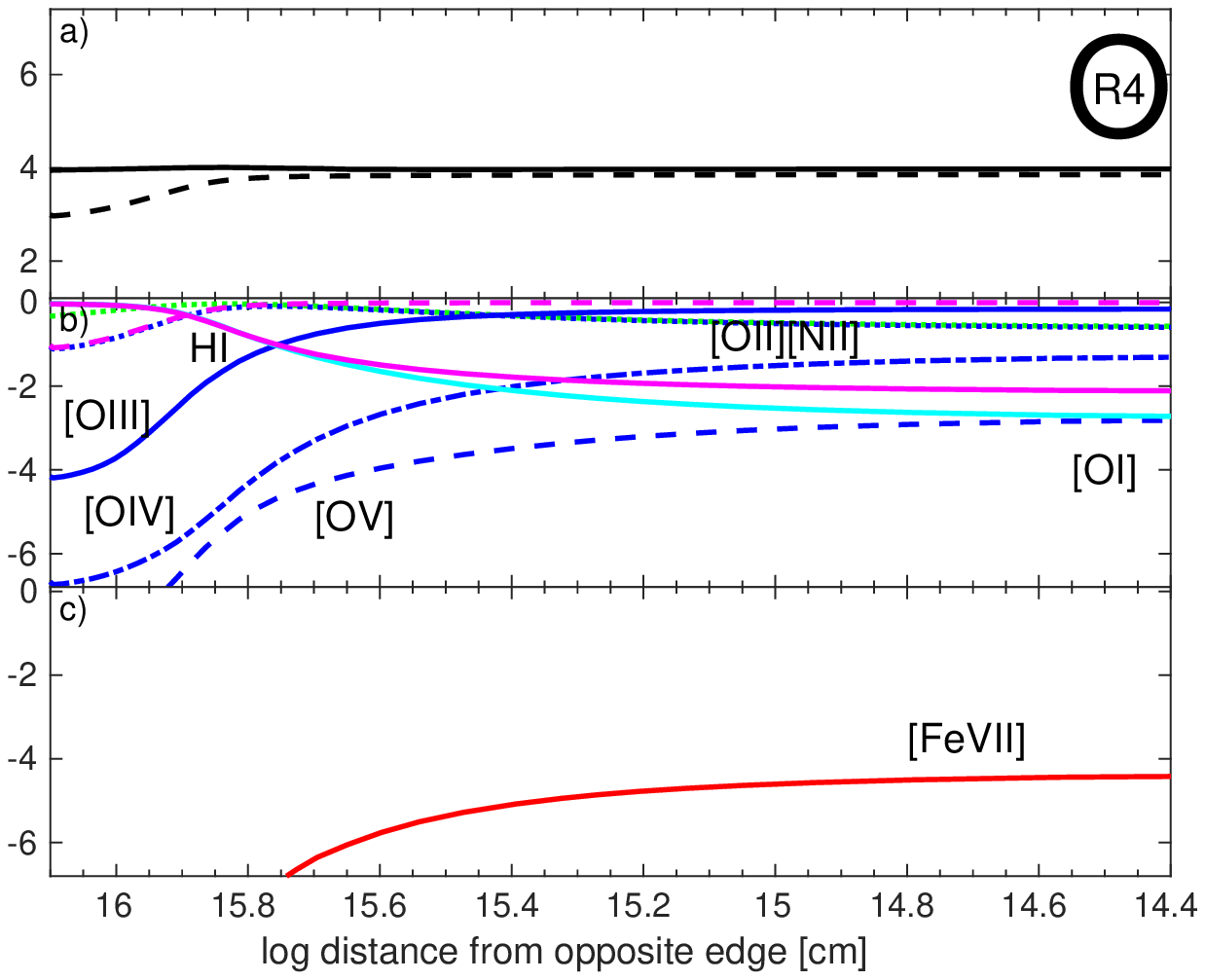}
\includegraphics[width=8.8cm]{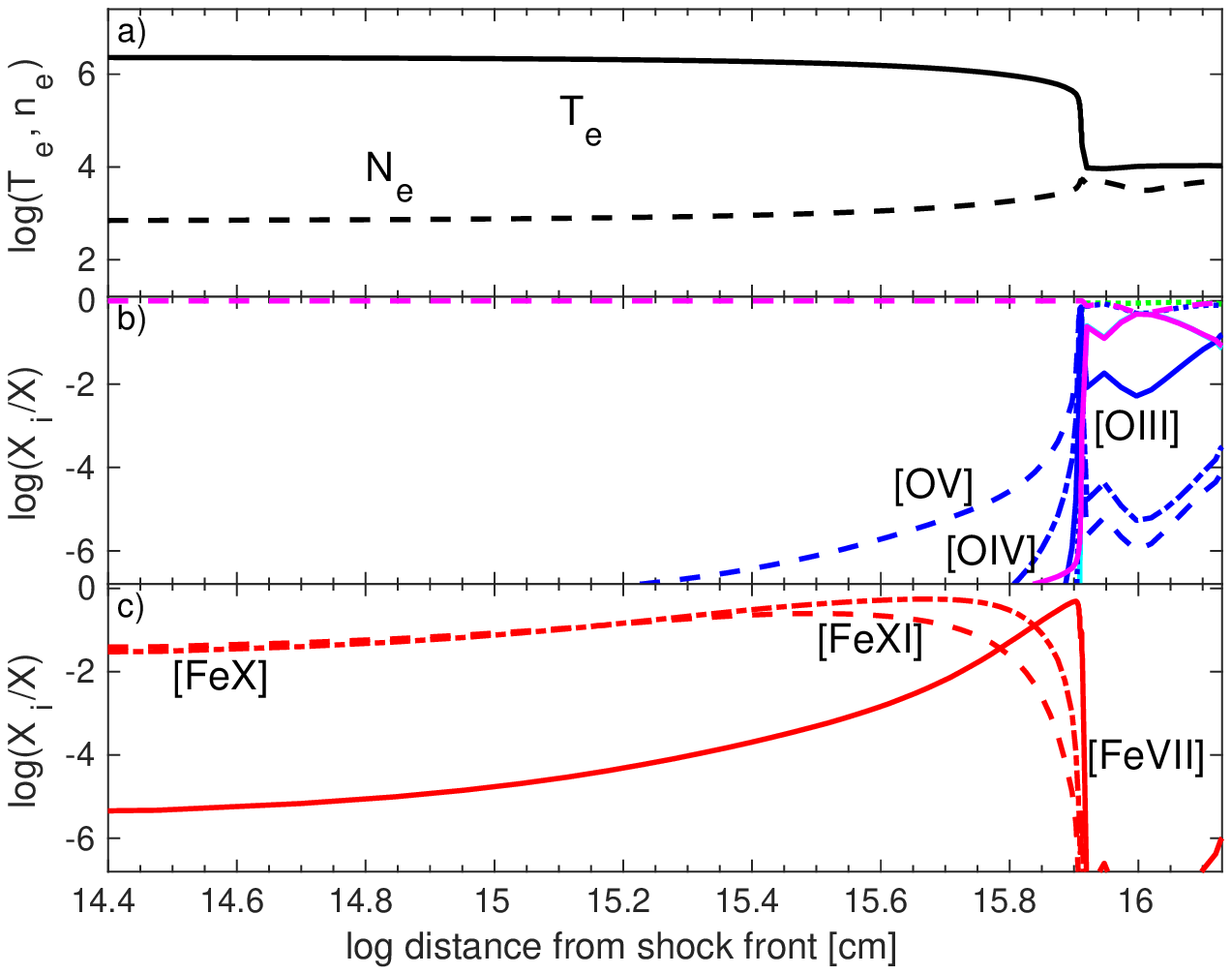}
\includegraphics[width=8.8cm]{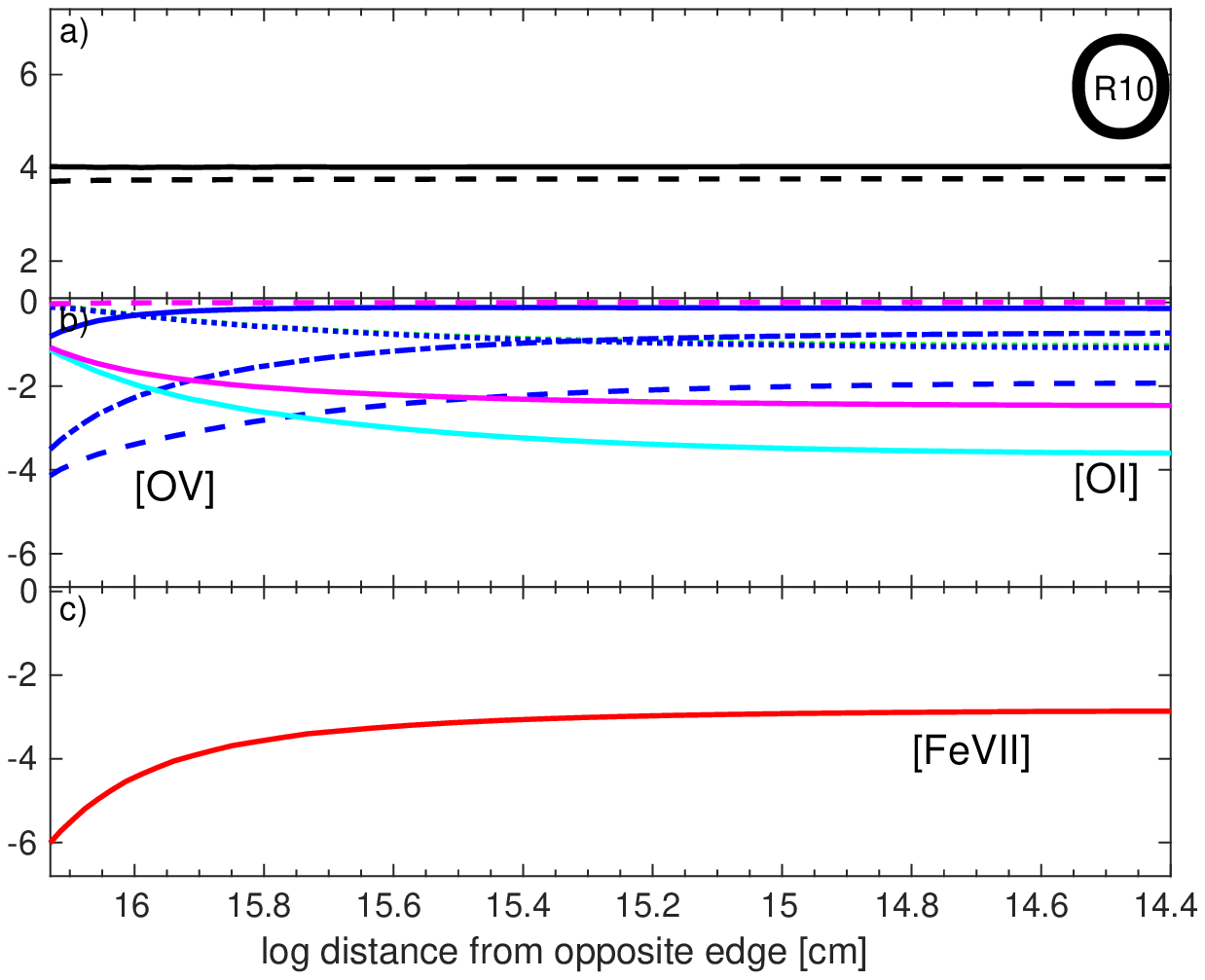}
	\caption{ Same as in Fig. \ref{fig_abundancia1} for regions R4 (top panels) and R10 (bottom panels).}
 \label{fig_abundancia2}
\end{figure*}

To better understand the modelling results, the profiles of  
 electron temperatures and densities (\Te\ and \Ne, respectively) and   of the
fractional abundances of the most significant ions  calculated throughout the clouds are 
displayed in Fig.~\ref{fig_abundancia1} for the R6 (top) and R7 (bottom) clouds. They are representative of clouds with  the minimum and maximum geometrical width, respectively. In addition, results for R4 (top) and R10 (bottom) clouds are in Fig.~\ref{fig_abundancia2}. They represent clouds in regions away from the SE-NW axis (R4) and far from the active centre (R10).
The corresponding %emission lines
atomic species within the clouds are marked in the figures.

 Tables \ref{tab:3} - \ref{tab:4} list the calculated percent contributions of SD and RD clouds to the emitted lines for 
each region. The most evident finding is that the [\ion{Fe}{vii}], [\ion{Fe}{x}], and [\ion{Fe}{xi}] lines are emitted from SD gas. In contrast, the RD clouds dominates for low- and mid-ionisation lines in most regions.

In Figs.~\ref{fig_abundancia1} and~\ref{fig_abundancia2} the clouds are divided into two  contiguous half regions on logarithmic scale in
order to show separately  the relative trend of the different lines at distances from the shock front and from the side
illuminated by AGN radiation.
For an outflowing gas the shock front is on the left of the left panel, opposite to the edge reached by the AGN radiation on the right of the right panel. The gas heated to  
temperatures  $\geq$ 10$^6$ K at the SD edge recombines slowly downstream  following 
the cooling rate.

The FWHM of the observed CLs indicate that the gas velocity is relatively high ($\geq 400$ \kms, see Fig.~\ref{fig:cinematica_ic5063}) in the different regions. As a consequence, the clouds are  possibly fragmented  by turbulence at the shock front \citep{contini_2011}.
In order to reproduce the [\ion{Fe}{x}] emission, which is observed  in the spectra of the R3, R6 and R7 regions, we will follow the 
method presented by \citet{fonseca-faria_2021}, i.e. we consider that cloud fragments that are not illuminated by radiation from the AGN also contribute to the spectra. The spectra from these SD fragments  (which we will call just “fragments” hereafter) are   summed up in pluri-cloud  models to the RD spectra, which account for both radiation and shocks.
The  fragments  are matter-bounded.  
Only the high ionisation-level (coronal) lines  emitted from the fragments are strong because they  come from the downstream region close to the shock front, where the temperatures are high. Low ionization-level lines are weak or null because the recombination zone within the cloud is cut off. Therefore, in pluri-cloud models only the coronal lines are affected by the fragments.
For regions R6 and R7, the  fragments have a very small geometrical thickness. R6 radiation dominated clouds are also matter-bounded.
The presence of clouds with different geometrical widths are  due to fragmentation in a turbulent regime created by shocks. It is possible that the fragments do not appear in regions far from the AGN.  So, even if the RD clouds in regions such as  R4 and R10 show  trends of \Ne, \Te\ and of the fractional abundances of the ions similar to those for R6, R7 and R3,
the coronal lines may have different intensities because the  latter are  dominated by 
the contribution from the SD fragments.

Top panels of Fig. \ref{fig_abundancia1} show that the RD region within the R6 clouds  is reduced  because the shock dominates. Moreover, the region of fully  recombined gas, where [\ion{O}{i}] stems from, disappeared.
On the other hand, the RD region  is very  extended in the R7 clouds (bottom panels). The geometrical thickness   $D$
which also affects the temperature distribution within  the clouds covers a rather large range (0.003-0.012 pc). 
$D$ bias the line ratios  as well as the relative SD/RD  gas emission.
The large range of $D$ is most probably  a consequence of the high fragmentation due to turbulence  near the shock front.

The  \Hb\ flux calculated at the nebula (\Hb$_c$) exceeds by many orders of magnitude  the flux observed at Earth (\Hb$_o$)  in all regions R1-R10.  
From their ratios \Hb$_c$/\Hb$_o$ = (d/r$_{c}$)$^2$, where d=49.3~Mpc is the distance  of IC5063 to Earth, we calculate the distance r$_{c}$  of 
each region to the AGN. Here, R6 acts as the  active centre.

As an additional test of consistency of the modelling, we compare the observed and calculated distances r$_o$ and r$_c$, respectively,  in the R1-R10 regions. The former distances were measured from the AGN in  Fig. \ref{fig:ratios_to_hbeta}.
Rows 23 and 25 in Tables \ref{tab:m1} and \ref{tab:m2} list r$_c$ and r$_o$, respectively.
For R4, R5, R8 and R9  the fit is acceptable within a
$\sim$ 10 percent error.  For R1 and R10 the error  is larger because  these regions are more extended and they are located at large distances from R6.   At distances of e.g. 2.2 kpc the error is  $\sim$ 30 percent.
For the clouds in region R6 we adopt a maximum not null distance from the AGN r$_o$ = 0.1 kpc.
To recover the  r$_o$ values  for  r$_c$  in the R2, R3, R6 and R7 regions
we introduce  the  filling  factors \ff=0.6, 0.13, 0.004, and 0.21, respectively.
Therefore, the filling value for each region was adjusted so that the values of r$_o$ match the values of r$_c$. It should be notice that in the shock scenario, the leading edge of the shock compresses pre-shock gas into a very small volume, producing a filling factor that is significantly lower than that of the circumnuclear ISM. The filling factor in such pre-shock gas is already low in the NLR of typical AGNs \citep{maksym+19}.
For the other regions \ff$\sim$1 (bottom row in Tables \ref{tab:m1}-\ref{tab:m2} ). The
\ff\ factor decreases with decreasing  distance from the  AGN  along the
SW direction  and it  drops to a minimum in the region adjacent to the   active centre.  Most probably the turbulent activity close to the AGN  breaks and destroys the clouds.

\subsection{Discussion}

The results of modelling presented in the last section showed that the bulk of the recombination lines (e.g. \Hb), the forbidden  intermediate  
(e.g. [\ion{O}{iii}]), and the low level lines
(e.g. [\ion{N}{ii}]) come from the RD zone, while high-ionisation level lines ([\ion{Fe}{vii}] and [\ion{Fe}{x}]) form in  the SD zone, 
which is heated to 
high temperatures  through the shock front and downstream. Therefore, the effect of  shocks is necessary to explain 
the relatively high [\ion{Fe}{vii}]/\Hb\ and [\ion{Fe}{x}]/\Hb\ line ratios observed in IC\,5063.  This is in agreement with previous works \citep{sharp_2010, mingozzi_2019, venturi_2021, HOLDEN_2022}, which  found that radiation from the AGN is the main source of ionisation for low- and mid-ionisation lines. Here, using a different approach, we reached to a similar conclusion. Our modelling points out that extended coronal emission is an unambiguous signature of shocks.

We interpret the nuclear and the two secondary peaks of high-ionisation gas as shock-driven emission produced by the interaction of the jet with the ISM due to the following reasons. First, at the nucleus, because of the projected galaxy scale (1~arcsec = 239~pc), we observed predominantly the gas emission coming from the region outside the torus and extended up to a few hundred parsecs. We have already seen that the coronal gas is highly turbulent and kinematically perturbed there. Very likely, the part of the CLR photoionised by the radiation from the central source is restricted to the central few tens of parsecs, as shown by~\citet{ardila_2017} in NGC\,4388. They showed that the AGN alone can produce strong coronal emission in the central few parsecs. Moreover, the extended coronal emission attributed to the AGN reaches a few tens of parsecs at the most. That scenario can be also applied to IC~5063 because both sources have a very similar bolometric luminosity. However, in IC~5063 the jet propagates very close to the galaxy plane, increasing the jet-ISM interaction. Second, the off-nuclear peaks of coronal emission coincides spatially with the position of the radio-lobes. Third, the coronal gas kinematics is strongly perturbed at the radio-lobes and with an enhance of the gas density (see Sect.~\ref{sec:physical_cond}). Fourth, the coronal gas is extended the most to the NW, where the radio-lobe is the brightest. Last but not the least, at distances equivalent to that of the radio-lobes but in the perpendicular direction to the radio-jet axis, no evidence of coronal emission is observed. However, low-ionisation gas fills up the inter-cone region (see Section~\ref{sec:physical_cond}).

From all the results gathered above, we conclude that the detection of extended coronal gas, at several hundreds of parsecs away from the AGN, is a natural consequence of the presence of shocks. In IC\,5063, the alignment between the radio-jet and the high-ionisation gas strongly suggest that the interaction between the jet and the ISM gas is the source of such shocks. The jet traces the kinetic feedback of the highest ionised component at optical and NIR wavelengths. Photoionisation models combining the effects of shocks and radiation from the active centre strongly support this scenario. Low- and mid-ionisation lines are primarily AGN-driven, as found by previous works, but we do not discard that shocks may also contribute to the observed emission line flux.

\section{Conclusions}
\label{sec:final}

We studied the highest ionised portion of the gas in the Seyfert~2 galaxy IC~5063 by means of VLT/MUSE IFU data in the optical region. To this purpose, we analysed the nuclear and extended highest-ionised emission in IC\,5063 using the coronal lines [\ion{Fe}{vii}]~$\lambda$6087 and [\ion{Fe}{x}]~$\lambda$6375. Moreover, we report the first detection of [\ion{S}{xii}]~$\lambda$7611 and [\ion{Fe}{xi}]~$\lambda$7892 in IC\,5063. Their emission region, though, is spatially unresolved and centred at the AGN. We detected extended coronal emission in the [\ion{Fe}{vii}]~$\lambda$6087 and [\ion{Fe}{x}]~$\lambda$6375 lines along the radio jet axis, with a total extension of $\sim$2~kpc from the SE to the NW edges of the coronal emission region. In addition, two secondary peaks of emission were found at 400~pc and 500~pc, SE and NW, respectively, from the AGN, coinciding with the position of the two radio lobes observed in this source. The largest extension of the coronal gas is located at 1193$\pm$39~pc NW of the AGN. We interpret the two secondary peaks of high-ionisation gas as shock-driven emission produced by the interaction of the jet
with the ISM. The evidence gathered from IC\,5063 and other AGNs, allowed to conclude that the detection of
extended coronal gas is a clear signature of the presence of a radio
jet strongly interacting with the ISM. They are the clearest evidence
of most energetic component of the kinetic feedback in a yet radio-weak AGN.

Additional support to the above scenario is obtained from a detailed analysis of the gas kinematics. We found that the coronal gas is decoupled from the disc rotation. Indeed, velocities in excess of up to 300~\kms\ relative relation to the gas rotation mapped by H$\alpha$ are identified and coincident with the position of
the radio lobes. At these locations, high FWHM values are identified,
of up to 500~\kms\, which coincide with high flux values of [\ion{Fe}{vii}]. These results indicate the presence of an expanding
and highly turbulent gas. Channel maps reveal broad blue-asymmetric profiles, with velocities reaching -600~\kms\ relative to the systemic velocity of the galaxy. To the SE, split line profiles are identified and associated to expanding gas. Velocities in excess of
350~\kms\ relative the systemic are found in this region. Furthermore, the fact that the highest temperature values are also co-spatial to regions of greater turbulence suggests that the extended coronal emission is likely driven by shocks produced by the interaction of the radio jet and the ionisation cone gas.  

The above scenario is further tested by means of photoionsation models that combine the effects of shocks and the radiation from a central source. Ten representative regions of the nuclear and circumnuclear environment across the FoV covered by MUSE were studied. We found that the bulk of the low- to medium-ionisation lines are produced by radiation dominated clouds, illuminated by the AGN continuum radiation flux. In contrast, most of the observed high ionisation lines ([\ion{Fe}{vii}], [\ion{Fe}{x}], and [\ion{Fe}{xi}]) are formed in shock dominated clouds. The shock velocities necessary to reproduced the  observed spectra are in agreement with those measured from the gas kinematics. We verify that the shocks are necessary to explain the relatively high line ratios [\ion{Fe}{vii}]/\Hb\ and [\ion{Fe}{x}]/\Hb\ observed in the nucleus and in the extended region.
Furthermore, the high values of density ($>10^2$~cm$^{-3}$) and temperature  (reaching 20000~K), associated to regions of greater turbulence and enhanced coronal line emission, corroborate the scenario of shocks produced by the interaction between the radio jet and the ISM.

\section*{Acknowledgements}
 
We thank to the Anonymous referee for useful comments and suggestions to improve this manuscript. ARA acknowledges Conselho Nacional de Desenvolvimento Cient\'{\i}fico e Tecnol\'ogico (CNPq) for partial support to this work through grant 312036/2019-1.

%%%%%%%%%%%%%%%%%%%%%%%%%%%%%%%%%%%%%%%%%%%%%%%%%%
\section*{Data Availability}

The  data  underlying  this  article  are  available  in  the  European Southern Observatory  archive.  The  data  can  be  obtained  in  raw quality through the MUSE raw data query form (\url{http://archive.eso.org/wdb/wdb/eso/muse/form}). Science quality data can be obtained from the Spectral data products query form (\url{http://archive.eso.org/wdb/wdb/adp/phase3spectral/form?collectionname=MUSE}).


\bibliographystyle{mnras}
\bibliography{referencia} % if your bibtex file is called example.bib

%\begin{thebibliography}{99}
%\end{thebibliography}

%%%%%%%%%%%%%%%%%%%%%%%%%%%%%%%%%%%%%%%%%%%%%%%%%%

%%%%%%%%%%%%%%%%% APPENDICES %%%%%%%%%%%%%%%%%%%%%

%\appendix



%%%%%%%%%%%%%%%%%%%%%%%%%%%%%%%%%%%%%%%%%%%%%%%%%%

% Don't change these lines
\bsp	% typesetting comment
\label{lastpage}
\end{document}